\documentclass[12pt]{article}
\usepackage{amsmath,amssymb}
\usepackage{graphicx,psfrag,epsf}
\usepackage{enumerate}
\usepackage{natbib}
\usepackage{url} 

\usepackage{multirow,color}

\newcommand{\blind}{0}

\DeclareMathOperator{\esssup}{ess\,sup}
\newtheorem{prop}{Proposition}
\newtheorem{lemma}{Lemma}
\newtheorem{theorem}{Theorem}
\newcommand{\on}{\operatorname}
\newcommand{\deq}{\,{\buildrel d \over =}\,}

\begin{document}

\def\spacingset#1{\renewcommand{\baselinestretch}%
{#1}\small\normalsize} \spacingset{1}


\if0\blind
{
  \title{\bf Efficient spatial modelling using the SPDE approach with bivariate splines}
  \author{Xiaoyu Liu \\
    Department of Statistical Science, University College London\\
    Serge Guillas \\
    Department of Statistical Science, University College London\\
    and \\
    Ming-Jun Lai \\
    Department of Mathematics, University of Georgia}
  \maketitle
} \fi

\if1\blind
{
  \bigskip
  \bigskip
  \bigskip
  \begin{center}
    {\LARGE\bf Efficient spatial modelling using the SPDE approach with bivariate splines}
\end{center}
  \medskip
} \fi

\bigskip
\begin{abstract}
Gaussian fields (GFs) are frequently used in spatial statistics for their versatility. The associated computational cost can be a bottleneck, especially in realistic applications. It has been shown that computational efficiency can be gained by doing the computations using Gaussian Markov random fields (GMRFs) as the GFs can be seen as weak solutions to corresponding stochastic partial differential equations (SPDEs) using piecewise linear finite elements. We introduce a new class of representations of GFs with bivariate splines instead of finite elements. This allows an easier implementation of piecewise polynomial representations of various degrees. It leads to GMRFs that can be inferred efficiently and can be easily extended to non-stationary fields. The solutions approximated with higher order bivariate splines converge faster, hence the computational cost can be alleviated. Numerical simulations using both real and simulated data also demonstrate that our framework increases the flexibility and efficiency.
\end{abstract}

\noindent%
\vfill

\newpage
\section{Introduction}
\label{sec:intro}
Gaussian fields (GFs) are at the core of spatial statistics, especially in the class of structured additive regression models, named latent Gaussian models, which are flexible and extensively used \citep{cressie1993statistics, banerjee2004hierarchical, diggle2007model}. However, when making statistical inference, it is usually needed to evaluate the likelihood function or the latent Gaussian field distribution, for which we need to make computations on dense matrices, e.g. the covariance matrix $\boldsymbol{\Sigma}(\boldsymbol{\theta})$, typically of order $\mathcal{O}(n^3)$ where $n$ is the dimension of $\boldsymbol{\Sigma}(\boldsymbol{\theta})$. \cite{rue2009approximate} overcome this computational hurdle. They approximate Bayesian inference in latent Gaussian models by assuming that the latent field is Gaussian Markov random field (GMRF). With only a few hyperparameters, integrated nested Laplace approximations (INLA) produce faster inference than simulation based approaches such as MCMC.  To take advantage of the computational efficiency of GMRF, \cite{lindgren2011explicit} constructed an explicit link between GFs and GMRFs. They considered the GFs with Mat\'ern covariance function,
\begin{equation}\label{matern}
r(\mathbf{u},\mathbf{v})=\frac{\sigma^2}{2^{\nu-1}\Gamma(\nu)}(\kappa\|\mathbf{v}-\mathbf{u}\|)^\nu K_{\nu}(\kappa\|\mathbf{v}-\mathbf{u}\|),
\end{equation}
where $\|\mathbf{v}-\mathbf{u}\|$ is the Euclidean distance between two locations $\mathbf{u}$ and $\mathbf{v}\in\mathbb{R}^D$, $K_{\nu}$ is the modified Bessel function of the second kind and order $\nu > 0$, $\kappa > 0$ controls the nominal correlation range through $\rho=\sqrt{8\nu}/\kappa$ corresponding to correlations near $0.1$ at the Euclidean distance $\rho$, and $\sigma^2$ is the marginal variance. The integer value of  $\nu$ determines the mean-square differentiability of the underlying process. 
They noticed that a Gaussian field $x(\mathbf{u})$ with Mat\'ern covariance $(\ref{matern})$ is a solution to the linear fractional SPDE
\begin{equation}\label{spde}
(\kappa^2-\Delta)^{\alpha/2} (\tau x(\mathbf{u}))=W(\mathbf{u}),~~~\mathbf{u}\in \mathbb{R}^D, ~~\alpha=\nu+d/2, ~~\kappa > 0, ~~\nu > 0,
\end{equation}
where the innovation process $W$ is spatial Gaussian white noise with unit variance \citep{whittle1954stationary,whittle1963stochastic}, $\Delta = \sum_{i=1}^d\frac{\partial^2}{\partial x_i^2}$ is the Laplacian operator, and $\tau$ controls the marginal variance through the relationship
 $$\tau^2=\frac{\Gamma(\nu)}{\Gamma(\nu+d/2)(4\pi)^{d/2}\kappa^{2\nu}\sigma^2}.$$

Therefore to find the GF $x(\mathbf{u})$ with covariance function $(\ref{matern})$ is to find the solution to $(\ref{spde})$. Denote the inner product of two functions $f$ and $g$ on $\mathbb{R}^D$ as
$\langle f,g \rangle = \int_{\mathbb{R}^D} f(\mathbf{u})g(\mathbf{u})\on{d}\mathbf{u},$
then the weak stochastic solution to SPDE $(\ref{spde})$ can be found by requiring that
\begin{equation}\label{spde_solver}
\langle \phi, (\kappa^2-\Delta)^{\alpha/2}\tau x \rangle\,{\buildrel \on{d} \over =}\,\langle \phi, W \rangle,
\end{equation}
for suitable functions $\phi(\mathbf{u})$, where `$\,{\buildrel \on{d} \over =}\,$' denotes equality in distribution \citep{walsh1986introduction}. Then \cite{lindgren2011explicit} constructed a finite element representation \citep{brenner2008mathematical} of the Gaussian random field over an unstructured triangulation of the form
\begin{equation}\label{fem}
x_h(\mathbf{u})=\sum_{k=1}^n w_k\psi_k(\mathbf{u}),
\end{equation}
where $\{\psi_k\}_{k=1}^n$ are piecewise linear basis functions. They showed that the Gaussian weights $\{w_k\}_{k=1}^n$ are GMRFs when $\alpha=1$, and can be approximated with GMRFs when $\alpha\geq 2$. Therefore the computations for GFs can be carried out using GMRFs and the computational efficiency can be improved dramatically. This work is closely related to the spatial spline regression models by \cite{sangalli2013spatial} where a spatial surface is approximated with finite elements. Another recent related work is \cite{nychka2014multi}, where the authors proposed a representation of a random field using multi-resolution radial basis functions on a regular grid. They also assumed that the coefficients associated with the basis functions to be distributed according to a GMRF to speed up the computation.

It is stated in \cite{lindgren2011explicit} and \cite{simpson2012think} that the convergence rate of a finite element approximation to the full solution to the SPDE $(\ref{spde})$ is of order $\mathcal{O}(h^2)$ where $h$ is the length of longest edge in the triangulation. Hence the convergence can be achieved by refinement of the underlying triangulation which is usually called the $h$-version finite elements. An alternative is to increase the approximation order over any fixed triangulation with higher degree polynomials over each triangle, which is called the $p$-version finite elements (the degree of polynomials is usually denoted by $p$). It has been illustrated that the convergence rate of the $p$-version cannot be worse than the $h$-version in most cases \citep{babuska1981p}. To increase the approximation order over each triangle, multivariate splines over triangulations can be employed instead of conventional finite elements. This provides a flexible and easy construction of splines with piecewise polynomials of various degrees and smoothness. Basic concepts and theories of multivariate splines can be found in the monograph by \cite{lai2007spline}. Multivariate splines have been shown to be more efficient and flexible than conventional finite element method in data fitting problems and solving PDEs, see  \cite{awanou2005multivariate}. It has been applied in spatial statistics. For example, \cite{guillas2010bivariate} introduced a spatial data analysis model with bivariate splines by penalizing the roughness with a partial differential operator; this has been demonstrated to be more efficient and accurate than thin-place splines in the application of ozone concentration forecasting \citep{ettinger2012bivariate}. In this paper, we introduce bivariate splines to represent the GFs on $\mathbb{R}^2$ and show its advantages over the piecewise linear finite elements used in \cite{lindgren2011explicit}. Within our framework of the SPDE approach using bivariate splines, it is allowed to choose piecewise polynomial representations of arbitrary degrees to adapt to the various data structures and features, and make the inference more computationally efficient.

The paper is structured as follows. In Section \ref{sec:bs_spde} some basics of bivariate splines in the Bernstein form (B-form) are reviewed first. Then we show how to link the GFs with GMRFs within the framework of bivariate splines, establish the theoretical properties of the bivariate spline approximations and discuss extensions to non-stationary fields. In Section \ref{sec:numerical}, we conduct several numerical simulations to illustrate our method and compare with the approach of \cite{lindgren2011explicit} on both real and simulated data sets. Section \ref{sec:discuss} consists of conclusion and discussion. Proofs are in the Appendix.

\section{SPDE approach using bivariate splines}
\label{sec:bs_spde}
\subsection{B-form bivariate splines}
\label{subsec:bs}
Let $\boldsymbol{\Delta}$ be a triangulation of a bounded domain $\Omega \subset \mathbb{R}^2$. We consider the continuous spline spaces
$$S_d^0(\boldsymbol{\Delta}) = \{s\in C^0(\Omega), s|_T\in \mathcal{P}_d, \forall T\in \boldsymbol{\Delta}\},$$ where $\mathcal{P}_d$ is the space of bivariate polynomials of degree $d\geq 1$, $C^0(\Omega)$ is the space of all continuous functions on $\Omega$. For any $d\geq 1$, the spline space $S_d^0(\boldsymbol{\Delta})$ contains all possible continuous spline functions which are bivariate polynomials of degree $d$ over each triangle $T\in \boldsymbol{\Delta}$. We apply the B-form representation of splines in $S_d^0(\boldsymbol{\Delta})$ proposed by \cite{awanou2005multivariate} in this paper, which allows an easy construction of locally supported basis functions and associated calculations. In this section we only give a brief introduction to the bivariate splines and some details are relegated to Appendix \ref{app:bs}. For more complete and in-depth explanations, see \cite{lai2007spline}.

Let $T=\left\langle \mathbf{v}_1,\mathbf{v}_2,\mathbf{v}_3 \right\rangle$ be a non-degenerate (i.e. with non-zero area) triangle with vertices $\mathbf{v}_1=(x_1,y_1)$, $\mathbf{v}_2=(x_2,y_2)$ and $\mathbf{v}_3=(x_3,y_3)$. Then every point $\mathbf{v}=(x,y)\in \mathbb{R}^2$ has a unique representation in the form
\begin{equation}\label{barycenter}
\mathbf{v}=b_1 \mathbf{v}_1+b_2 \mathbf{v}_2+b_3 \mathbf{v}_3,
\end{equation}
with $b_1+b_2+b_3=1$, where $b_1$, $b_2$ and $b_3$ are named the barycentric coordinates of the point $\mathbf{v}=(x,y)$ relative to the triangle $T$. The polynomials
\begin{equation}\label{bpoly}
B_{ijk}^{T,d}(\mathbf{v}) = \frac{d}{i!j!k!}b_1^i b_2^j b_3^k,
\end{equation}
are called the Bernstein polynomials of degree $d$ relative to triangle $T$. Then for each spline function $s\in S_d^0(\boldsymbol{\Delta})$, we can write
$$s|_T=\sum_{i+j+k=d}c_{ijk}^TB_{ijk}^{T,d},~~~T\in \boldsymbol{\Delta},$$
where the coefficients $\mathbf{c} = \{c_{ijk}^T, i+j+k=d, T\in\boldsymbol{\Delta}\}$ are called B-coefficients of $s$.

\subsection{SPDE modelling with B-form bivariate splines}
\label{subsec:bs_spde}
Let $\{\psi_1, \psi_2,..., \psi_m\}$ be a set of locally supported basis functions of $S_d^0(\boldsymbol{\Delta})$ for any $d\geq 1$, where $m=\on{dim} S_d^0(\boldsymbol{\Delta})$ (see Appendix \ref{app:bs} and \cite{lai2007spline} for more details on these basis functions). For any $h=1,...,m$, the corresponding B-coefficients of $\psi_h$ are denoted by $\mathbf{c}_h$. Then the weak solution to the SPDE $(\ref{spde})$ in the spline space $S_d^0(\boldsymbol{\Delta})$ can be represented as
\begin{equation}\label{bs_x}
x_{\boldsymbol{\Delta}}(\mathbf{u})=\sum_{h=1}^m w_{h}\psi_{h}(\mathbf{u}).
\end{equation}

Remember that the weak solution to SPDE $(\ref{spde})$ can be found by requiring $(\ref{spde_solver})$ for any sensible test functions. However it is not possible to test all functions. Following \cite{lindgren2011explicit} we can choose a finite set of test functions. Specifically, we choose $\phi_h=(\kappa^2-\Delta)^{1/2}\psi_h$ for $\alpha=1$ leading to the least squares solution. For $\alpha=2$, we can choose either $\phi_h=\psi_h$ for any $d\geq 1$ or $\phi_h=(\kappa^2-\Delta)\psi_h$ for $d\geq 2$, leading to the Galerkin or least squares solution respectively. For $\alpha\geq 3$, if we let $\alpha=2$ on the left-hand side of the SPDE $(\ref{spde})$ then the right-hand side is a Gaussian process generated by the operator $(\kappa^2-\Delta)^{(\alpha-2)/2}$. Then we can choose $\phi_h=\psi_h$ for this innovative SPDE. Hence we get a recursive Galerkin solutions ending with either $\alpha=1$ or $2$. Therefore we have the main results as below.

\begin{theorem}\label{Q_theorem}
The vector of weights $\mathbf{w}=(w_h,h=1,...,m)'$ of bivariate spline representation for the solution to SPDE $(\ref{spde})$ defined in $(\ref{bs_x})$ is Gaussian with mean zero and the precision matrix $\mathbf{Q}_\alpha$ are given as follows:
\begin{itemize}
\item[(1)] for $\alpha=1$,
$$\mathbf{Q}_1=\tau^2(\kappa^2\mathbf{M}+\mathbf{K}),$$
\item[(2)] for $\alpha=2$,
$$\mathbf{Q}_2^{G}=\tau^2(\kappa^4\mathbf{M}+2\kappa^2\mathbf{K}+\mathbf{K}\mathbf{M}^{-1}\mathbf{K}),$$
$$\mathbf{Q}_2^{LS}=\tau^2(\kappa^4\mathbf{M}+2\kappa^2\mathbf{K}+\mathbf{R}),$$
where $\mathbf{Q}_2^{G}$ and $\mathbf{Q}_2^{LS}$ are the Galerkin and least squares solutions respectively,
\item[(3)] for $\alpha\geq 3$,
$$\mathbf{Q}_{\alpha}=\kappa^4\mathbf{Q}_{\alpha-2}+\kappa^2(\mathbf{Q}_{\alpha-2}\mathbf{M}^{-1}\mathbf{K}+\mathbf{K}\mathbf{M}^{-1}\mathbf{Q}_{\alpha-2})+\mathbf{K}\mathbf{M}^{-1}\mathbf{Q}_{\alpha-2}\mathbf{M}^{-1}\mathbf{K},$$
\end{itemize}
where
$$\mathbf{M}=\mathbf{C}'\mathbf{M_0}\mathbf{C},~~\mathbf{K}=\mathbf{C}'\mathbf{K_0}\mathbf{C},~~\mathbf{R}=\mathbf{C}'\mathbf{R_0}\mathbf{C},$$
and
$\mathbf{M_0}=\on{diag}(\mathbf{M}_T,T\in\boldsymbol{\Delta})$, $\mathbf{K_0}=\on{diag}(\mathbf{K}_T,T\in\boldsymbol{\Delta})$ and $\mathbf{R_0}=\on{diag}(\mathbf{R}_T,T\in\boldsymbol{\Delta})$ are block diagonal square matrix with square blocks
$$\mathbf{M}_T=\left[\int_T B_{ijk}^T(x,y) B_{\nu\mu\kappa}^T(x,y)dxdy\right]_{i+j+k=d}^{\nu+\mu+\kappa=d},$$
$$\mathbf{K}_T=\left[\int_T \nabla B_{ijk}^T(x,y) \nabla B_{\nu\mu\kappa}^T(x,y)dxdy\right]_{i+j+k=d}^{\nu+\mu+\kappa=d},$$
and
$$\mathbf{R}_T=\left[\int_T \Delta B_{ijk}^T(x,y) \Delta B_{\nu\mu\kappa}^T(x,y)dxdy\right]_{i+j+k=d}^{\nu+\mu+\kappa=d},$$
respectively and $\mathbf{C}=(\mathbf{c}_1,...,\mathbf{c}_m)$ whose $h$-th column is the B-coefficient vector of basis function $\psi_h$.
\end{theorem}

Since the basis functions $\{\psi_1, \psi_2,..., \psi_m\}$ are locally supported in $S^0_d(\boldsymbol{\Delta})$, the matrices $\mathbf{M}$, $\mathbf{K}$ and $\mathbf{R}$ are guaranteed to be sparse.
However in the Galerkin solution for $\alpha=2$ and recursive solutions for $\alpha\geq 3$, the inverse matrix $\mathbf{M}^{-1}$ can be not sparse, making the precision matrix dense. The mass lumping technique \citep{chen1985lumped} can be applied by replacing $\mathbf{M}$ with a diagonal matrix $\tilde{\mathbf{M}}$ whose elements are the sum of each row of $\mathbf{M}$, i.e. $\tilde{\mathbf{M}}_{ii}=\sum_j\mathbf{M}_{ij}$. Therefore the precision matrix is sparse and the underlying coefficients $\mathbf{w}$ are approximated with a GMRF. In the next sections we show the effect of such Markov approximation in both theory and application. Another point we want to mention is that piecewise linear finite elements are typical spline functions. The finite element representation in \cite{lindgren2011explicit} is just a bivariate spline in $S_1^0(\boldsymbol{\Delta})$.

\subsection{Approximation properties}
\label{subsec:approx}
Previously, we have constructed the bivariate spline approximation $x_{\boldsymbol{\Delta}}(\mathbf{u})$ to the true Gaussian random field or the solution to the SPDE $(\ref{spde})$. We explore some theoretical properties of our method in this section.

Define the Hilbert space $H^1$ associated with the differential operator $(\kappa^2-\Delta)$ to be the space of square integrable functions $f(x,y)$ for which $\|f\|_{H^1}^2=\kappa^2\int_{\Omega}f(x,y)^2dxdy+\int_\Omega\nabla f(x,y)\cdot \nabla f(x,y) dxdy$ is finite following \cite{lindgren2011explicit}. Approximation results for bivariate splines, e.g. Th. $5.19$ in \cite{lai2007spline}, show that the bivariate spline space $S_d^0(\boldsymbol{\Delta})$ for any $d\geq 1$ spanned by a finite set of basis functions $\{\psi_1,...,\psi_m\}$ is dense in $H^1$: for every $f\in H^1$, there is a sequence $\{f_m\}$, $f_m\in S_d^0(\boldsymbol{\Delta})$ such that $\lim_{m\to \infty}\|f-f_m\|_{H^1}=0$ where the limit scenario $m\to \infty$ corresponds to $|\boldsymbol{\Delta}|\to 0$ where $|\boldsymbol{\Delta}|$ is the length of the longest edge in the triangulation $\boldsymbol{\Delta}$. Using this fact, it follows directly from the Th. $3$-$4$ in Appendix C.2 of \cite{lindgren2011explicit} that, the bivariate spline approximation $x_{\boldsymbol{\Delta}}$ converges weakly to the weak solution to the SPDE. Note that the weak convergence of $x_{\boldsymbol{\Delta}}$ obtained for $\mathbf{Q}_2^{LS}$ cannot be derived directly but can be easily proved in the same fashion with just a few modifications. In addition, we can derive rates of convergence results. For example, when $\alpha=2$ we have the proposition below regarding to the Galerkin solutions.

\begin{prop}\label{convergence_rate}
Let $L=(\kappa^2-\Delta)$, $x_{\boldsymbol{\Delta}}(s)$ is the bivariate spline approximation of the random Gaussian field $x(s)$ in the spline space $S_d^0(\boldsymbol{\Delta})$, $d\geq 1$. Then for any $f\in H^1\cap W_2^{m+1}(\Omega)$ with $1\leq m \leq d$, where $W_2^{m+1}$ is a Sobolev space that is defined in Appendix \ref{app:prop1}, we have
$$\on{E}\left(\int_\Omega f(s)L(x(s)-x_{\boldsymbol{\Delta}}(s))ds\right)^2\leq K|\boldsymbol{\Delta}|^{m+1}|f|_{m+1,2,\Omega}$$
where $K$ is a constant, $|\boldsymbol{\Delta}|$ is the length of the longest triangle edge in the triangulation $\boldsymbol{\Delta}$ and $|f|_{m+1,2,\Omega}$ is defined in Appendix \ref{app:prop1}.
\end{prop}

It is clear that we are able to achieve a faster convergence rate by using bivariate splines with higher degree $d$. For example, when $d=3$ the convergence rate can be as high as $\mathcal{O}(|\boldsymbol{\Delta}|^4)$, which is two magnitude higher than $\mathcal{O}(|\boldsymbol{\Delta}|^2)$ in \cite{lindgren2011explicit}.

As we have mentioned, the matrix $\mathbf{M}$ in the Galerkin solutions is lumped by replacing $\mathbf{M}$ with a diagonal matrix $\tilde{\mathbf{M}}$ which yields a Markov approximation $\tilde{x}_{\boldsymbol{\Delta}}$ to the bivariate spline solution $x_{\boldsymbol{\Delta}}$.
Let $f$ and $g$ be test functions in $H^1$ and let $f_{\boldsymbol{\Delta}}$ and $g_{\boldsymbol{\Delta}}$ be their projections onto the bivariate spline space $S_d^0(\boldsymbol{\Delta})$ for any $d\geq 1$, with basis weights $\mathbf{w}_f$ and $\mathbf{w}_g$. Since the recursive algorithm for $\alpha\geq 3$ is based on $\alpha=2$ at each iteration, here we only investigate the effect of the Markov approximation on the Galerkin solutions for $\alpha=2$. When $\alpha=2$, the difference between the covariances for the Markov approximation $\tilde{x}_{\boldsymbol{\Delta}}$ and the bivariate spline solution $x_{\boldsymbol{\Delta}}$ is
$$\epsilon_{\boldsymbol{\Delta}}(f_{\boldsymbol{\Delta}},g_{\boldsymbol{\Delta}}) = \on{cov}(\langle f,L\tilde{x}_{\boldsymbol{\Delta}}\rangle_\Omega,\langle g,L\tilde{x}_{\boldsymbol{\Delta}}\rangle_\Omega)-\on{cov}(\langle f,Lx_{\boldsymbol{\Delta}}\rangle_\Omega,\langle g,Lx_{\boldsymbol{\Delta}}\rangle_\Omega)=\mathbf{w}_f'\tilde{\mathbf{M}}\mathbf{w}_g-\mathbf{w}_f'\mathbf{M}\mathbf{w}_g.$$
We have the following result showing that such a difference can be bounded.
\begin{prop}\label{Markov_approx}
For $f_{\boldsymbol{\Delta}}$, $g_{\boldsymbol{\Delta}}\in S_d^0(\boldsymbol{\Delta})$, we have
$$|\epsilon_{\boldsymbol{\Delta}}(f_{\boldsymbol{\Delta}},g_{\boldsymbol{\Delta}})|\leq K|\boldsymbol{\Delta}|^{2}
$$
where $K$ is a positive constant dependent on
$\|f_{\boldsymbol{\Delta}}\|_{2,2,\Omega},
\|g_{\boldsymbol{\Delta}}\|_{2,2,\Omega},
\|f_{\boldsymbol{\Delta}}\|_{\infty,\Omega},
\|g_{\boldsymbol{\Delta}}\|_{\infty,\Omega}$
(norms are defined in Appendix \ref{app:prop1}) and $|\boldsymbol{\Delta}|$ is the length of the longest triangle edge in
the triangulation $\boldsymbol{\Delta}$.
\end{prop}

We can see that the convergence rate of the bivariate spline representation in $S_d^0(\boldsymbol{\Delta})$ in Proposition $\ref{convergence_rate}$ may be decreased. But it is still at least $\mathcal{O}(|\boldsymbol{\Delta}|^2)$ in theory, which is as good as the linear finite elements representation. In fact, numerical simulations later illustrate that the approximation using bivariate splines with higher degree $d$ can be more efficient.

\subsection{Non-stationary fields}
\label{subsec:nonstationary}
\cite{lindgren2011explicit} showed that the SPDE $(\ref{spde})$ can be extended to a non-stationary version
\begin{equation}\label{spde_ns}
(\kappa^2(\mathbf{u})-\Delta)^{\alpha/2}(\tau(\mathbf{u})x(\mathbf{u})) = W(\mathbf{u}),
\end{equation}
where the parameters $\kappa^2$ and $\tau$ are not constants but depend on the location $\mathbf{u}$. The two parameters are assumed to vary slowly over the domain of $\mathbf{u}$ and have the general form of low dimensional representation
$$\log(\kappa^2(\mathbf{u}))=\sum_{j=1}^{n_{\kappa^2}}\theta_j^{(\kappa^2)}B_j^{(\kappa^2)}(\mathbf{u}),~~
\log(\tau(\mathbf{u}))=\sum_{j=1}^{n_{\tau}}\theta_j^{(\tau)}B_j^{(\tau)}(\mathbf{u}),$$
where the number of smooth basis functions $\{B_j^{(\cdot)}(\mathbf{u})\}$ for $\kappa^2(\mathbf{u})$ and $\tau(\mathbf{u})$ respectively, $n_{\kappa^2}$ and $n_{\tau}$, should not be large to guarantee computational efficiency. The inner product can be approximated with
$$\langle \psi_t,\kappa^2\psi_s\rangle\approx \kappa^2(\mathbf{u}_{s}^{\star})\langle \psi_t,\psi_s\rangle,$$
where $\mathbf{u}_{s}^{\star}$ is some point in the support of $\psi_s$ which can be chosen to be the domain point associated with the non-zero B-coefficients of $\psi_s$ (see Appendix \ref{app:bs} for more details about domain points and their relationship to $\psi_s$). 
Defining the diagonal matrices
$$\boldsymbol{\kappa}^2=\on{diag}(\kappa^2(\xi_h),h=1,...,m),~~\boldsymbol{\tau}=\on{diag}(\tau(\xi_h),h=1,...,m),$$
where $\xi_h$ is the domain point associated with the non-zero B-coefficients of basis function $\psi_h$ for $h=1,...,m$. It can be easily shown with minor modification of the proof of Theorem \ref{Q_theorem} that the weights $\mathbf{w}$ in the bivariate spline representation $(\ref{bs_x})$ can be approximated with GMRF as well. For example when $\alpha=2$, the precision matrix of $\mathbf{w}$ is
$$\mathbf{Q}_2^G(\boldsymbol{\kappa}^2,\boldsymbol{\tau})=\boldsymbol{\tau}(\boldsymbol{\kappa}^2\mathbf{M}\boldsymbol{\kappa}^2+2\boldsymbol{\kappa}^2\mathbf{K}+\mathbf{K}\mathbf{M}^{-1}\mathbf{K})\boldsymbol{\tau},$$
$$\mathbf{Q}_2^{LS}(\boldsymbol{\kappa}^2,\boldsymbol{\tau})=\boldsymbol{\tau}(\boldsymbol{\kappa}^2\mathbf{M}\boldsymbol{\kappa}^2+2\boldsymbol{\kappa}^2\mathbf{K}+\mathbf{R})\boldsymbol{\tau},$$
for Galerkin and least squares solutions respectively. As stated in \cite{lindgren2011explicit}, by assuming the parameters $\kappa^2$ and $\tau$ to be constant locally, the solution to the SPDE $(\ref{spde_ns})$ can still be interpreted as a Mat\'ern field over a local area and the associated global non-stationary field can be achieved by combining all the local Mat\'ern fields automatically via the SPDE.

\section{Numerical simulations}
\label{sec:numerical}
We conduct several numerical simulations to evaluate the performance of the SPDE approach with bivariate splines and compare with the linear finite element approach in \cite{lindgren2011explicit} in terms of spatial prediction. In all simulations over $\mathbb{R}^2$ we fix $\alpha=2$ which corresponds to the smoothness parameter $\nu=1$ in the Mat\'ern covariance function. The full Bayesian inference for the model is run in \texttt{R-inla (www.r-inla.org)} using the integrated nested Laplace approximation \citep{rue2009approximate}. For brevity, our proposed bivariate spline approximation in $S_d^0(\boldsymbol{\Delta})$ is denoted BS-SPDE with $d=1,2,...$ (BS-SPDE-G or BS-SPDE-LS for Galerkin or least squares solution respectively) while the linear finite element approximation is denoted LFE-SPDE.

\subsection{Comparison of LFE-SPDE and BS-SPDE}
\label{subsec:compare}
\subsubsection*{Study $1$}
\label{subsubsec:study1}
In this simulation, we compare the LFE-SPDE method and BS-SPDE-G or BS-SPDE-LS of degree $d\geq 2$ in data fitting and prediction for some common surfaces. Elevations of different surfaces are collected on a grid over square $[-2,2]\times[-2,2]$ that is equally spaced every $0.2$. We consider the specific surface to be an unknown random Gaussian field with Mat\'ern covariance which is also the solution to the stationary SPDE $(\ref{spde})$. The whole random field can be approximated using LFE-SPDE and Galerkin or least squares BS-SPDE respectively. Then elevations on another finer grid that is equally spaced every $0.01$ over square $[-2,2]\times[-2,2]$ can be predicted. The prediction accuracy for the whole surface can be measured with mean squared error $\on{MSE}=\sum_{i=1}^n (\hat{f}(\mathbf{u}_i)-f(\mathbf{u}_i))^2/n$, where $f(\mathbf{u}_i)$ is the true elevation on location $\mathbf{u}_i$ and $\hat{f}(\mathbf{u}_i)$ is the prediction using corresponding posterior means.

\begin{figure}[htbp!]
\centering
\includegraphics[width=15cm]{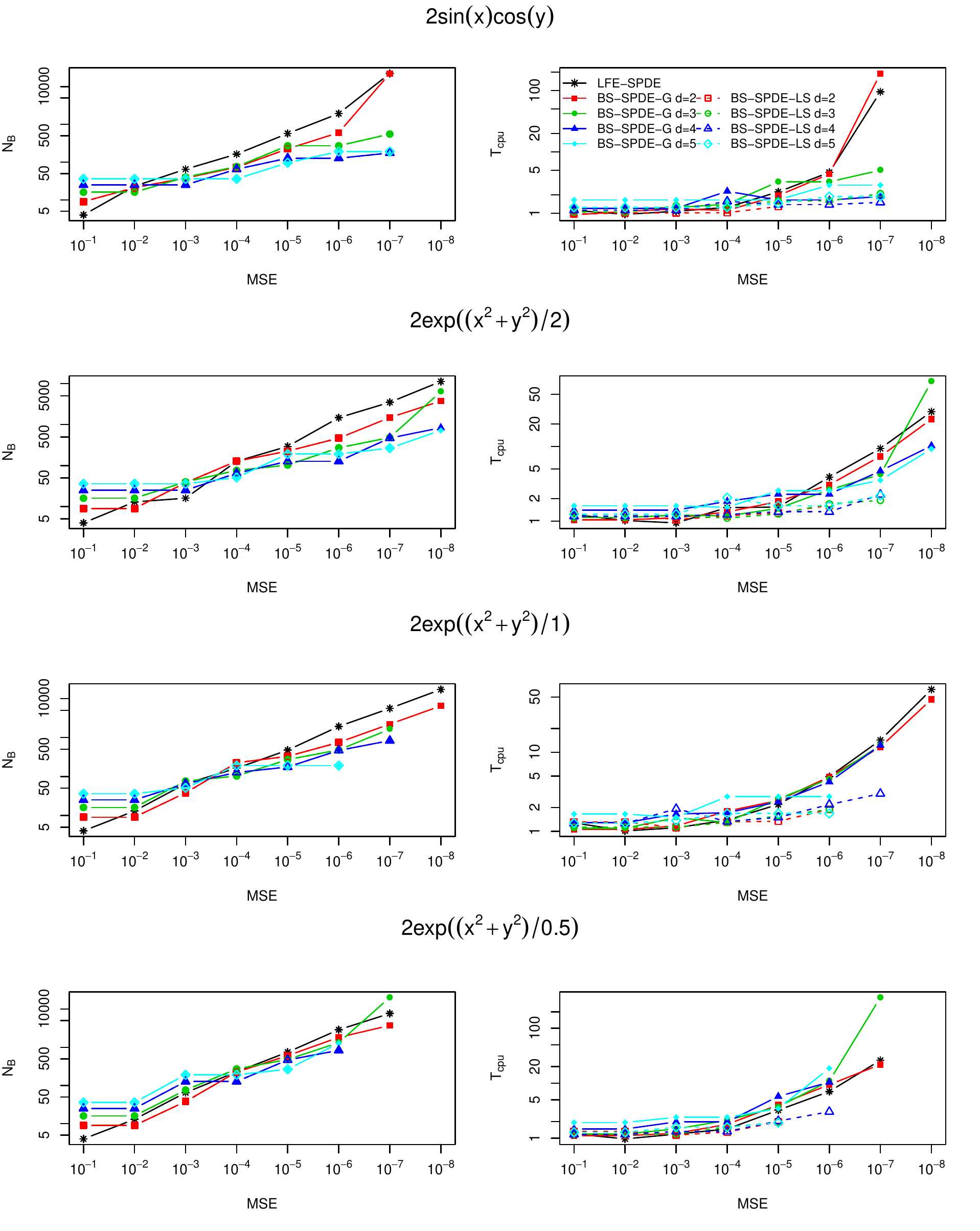}
\caption{Number of basis functions ($N_B$) and CPU time for \texttt{inla} ($T_{cpu}$ in seconds) required by LFE-SPDE, BS-SPDE-G and BS-SPDE-LS with $d=2,3,4,5$ respectively to reach specific MSE levels for different surfaces}
\label{known_suf_fig}
\end{figure}

We consider four different surfaces here including 
$2\sin(x)\cos(y)$ and $2\exp(-\frac{x^2+y^2}{s})$ with three different shape parameters $s=2$, $1$, $0.5$. As we have shown that both the finite element and bivariate spline approximations converge weakly to the full SPDE solution, we construct $35$ different meshes 
that have $2$, $3$, $6$, $13$, $19$, $28$, $53$, $96$, $112$, $148$, $212$, $279$, $342$, $390$, $444$, $520$, $705$, $874$, $1065$, $1368$, $1802$, $2416$, $2798$, $3176$, $3708$, $4428$, $5514$, $6696$, $8460$, $10958$, $15009$, $21832$, $26718$, $33776$, $43875$  triangles respectively to demonstrate the convergence (mesh size $|\boldsymbol{\Delta}|$ monotonically decreases roughly from $6.6$ to $0.026$). For each approach, the associated 
number of basis functions (denoted by $N_B$) and CPU time for \texttt{inla} (denoted by $T_{cpu}$ in seconds) are recorded when the corresponding MSEs reach levels of $10^{-1}$, $10^{-2}$, $10^{-3}$, $10^{-4}$, $10^{-5}$, $10^{-6}$, $10^{-7}$, $10^{-8}$. $N_B$ is also the dimension of corresponding precision matrix of the weights $\mathbf{w}$ and directly relates to the computational complexity in the associated calculations. For example the samples and likelihoods can be computed in $\mathcal{O}(N_B^{3/2})$ operations for two dimensional GMRFs. For comparison, the simulation stops when the number of basis functions of BS-SPDE with $d\geq 2$ exceeds the number of basis functions of LFE-SPDE using the densest mesh. The results are presented in Figure \ref{known_suf_fig} where the $y$-axes for $N_B$ and $T_{cpu}$ are taken on a logarithmic scale.

From Figure \ref{known_suf_fig} we can see that in general BS-SPDE with $d\geq 2$ can be more efficient than LFE-SPDE both in terms of number of basis functions and computing time needed to reach specific levels of MSE, especially those lower than $10^{-4}$. Note that in the left side of Figure \ref{known_suf_fig}, the dash lines for BS-SPDE-LS are invisible as they coincide with the solid lines for BS-SPDE-G while BS-SPDE-LS is more computationally efficient in general than BS-SPDE-G as shown in the right side when the associated precision matrices are of the same dimension. Specifically, for the surface $2\sin(x)\cos(y)$, to reach a specific level of MSE, especially high precision levels such as $10^{-6}$ or $10^{-7}$, BS-SPDE-G and BS-SPDE-LS with high degree $d\geq 3$ are more efficient since they require only less than $10\%$ of the basis functions and computing time required by LFE-SPDE. For the Gaussian surface $2\exp(-\frac{x^2+y^2}{2})$, BS-SPDE with $d\geq 2$ are generally much more efficient than LFE-SPDE for the MSE levels up to $10^{-7}$ with about $50\%$ gains in the computing time. But BS-SPDE-LS does not reach the MSE level $10^{-8}$ and BS-SPDE-G with $d=3$ takes more computing time than the others to reach the MSE level $10^{-8}$. For the next Gaussian shape surface $2\exp(-\frac{x^2+y^2}{1})$ which is steeper than the previous one, BS-SPDE with high degrees can be better than LFE-SPDE for the MSE levels around $10^{-4}$ to $10^{-6}$ but their efficiency is decreased to reach the higher precision levels $10^{-7}$ and $10^{-8}$. Only BS-SPDE-G with $d=2$ reaches the lowest MSE level $10^{-8}$. However, BS-SPDE-LS with $d=4$ reaches the low MSE level $10^{-7}$ within only $20\%$ of the computing time required by LEF-SPDE. For the last surface which is quite steep, even though neither BS-SPDE-G nor BS-SPDE-LS with $d\geq 2$ is more efficient than LFE-SPDE in most cases, BS-SPDE-G with $d=2$ is comparable with LFE-SPDE and reaches the high precision levels $10^{-6}$, $10^{-7}$ by requiring slightly less number of basis functions and similar time, and BS-SPDE-LS with $d=4$ is more efficient than the others to reach the MSE level $10^{-6}$.

From these results, we can conclude that BS-SPDE can be much more efficient in many cases especially when the high precision levels are desired and the target functions are smooth. For functions that are not that smooth, lower degree representations such as LFE-SPDE or BS-SPDE with $d=2$ might be more appropriate, which is also consistent with the general comments by \cite{babuska1981p}. Note that even for the last Gaussian shape surface which is much less smooth than the others, BS-SPDE-G with $d=2$ still can be comparable with LFE-SPDE; and we still obtain $50\%$ gains in the computing time using BS-SPDE-LS with $d=4$ if the MSE level $10^{-6}$ is desired.
\vskip 0.3cm

\subsubsection*{Study $2$}
\label{subsubsec:study2}
\begin{figure}[htbp!]
\centering
\includegraphics[width=7.5cm]{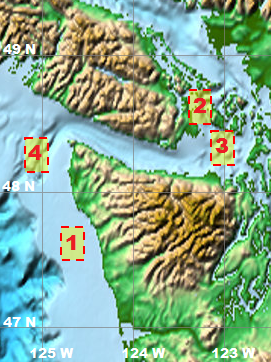}
\caption{Four regions extracted from ETOPO 1 around the Strait of Juan de Fuca from NGDC}
\label{cascadia_dataset}
\end{figure}

In this study, we compare LFE-SPDE and BS-SPDE in spatial estimation and prediction with real data sets that are extracted from the ETOPO$1$ Global Relief Model \citep{amante2009etopo1}, which is a $1$ arc-minute global relief model of Earth's surface that integrates land topography and ocean bathymetry. The data is available from National Geophysical Data Center (NGDC), USA. Four different regions around the the Strait of Juan de Fuca area are chosen for this study as shown in Figure \ref{cascadia_dataset}. In general, region $1$ covers relatively simple and gradual variations in the near shore seabed while the seabed in the other three regions is quite complicated.

\begin{figure}[htbp!]
\centering
\includegraphics[width=15cm]{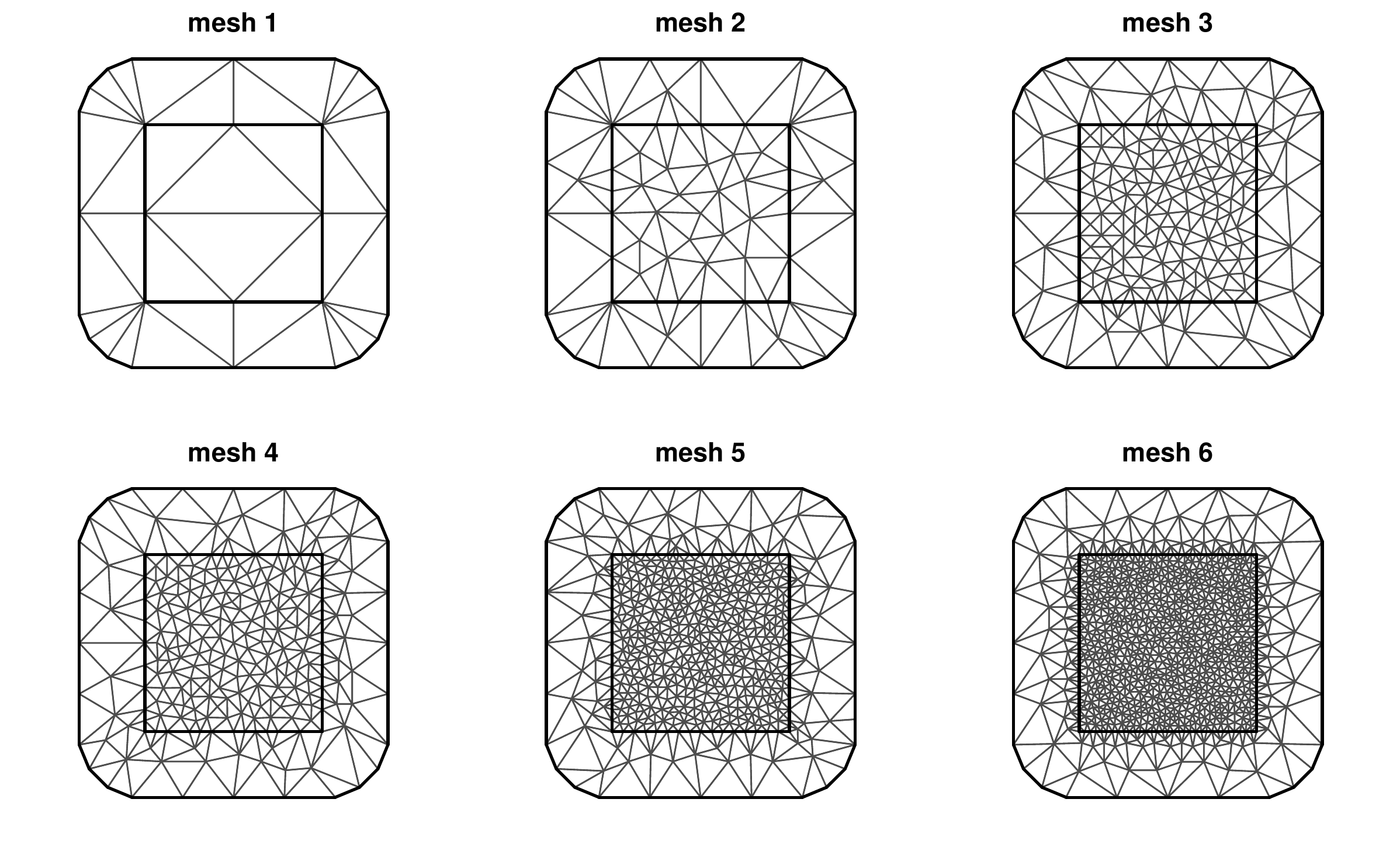}
\caption{Six meshes for Study $2$}
\label{cascadia_mesh}
\end{figure}

For comparison, both in-sample and out-of-sample predictive fit performance are explored using LFE-SPDE, BS-SPDE-G and BS-SPDE-LS with $d=2,3,4$ based on various meshes. We denote the observations by $y_1$, $y_2$, ..., $y_n$. As for the in-sample fit measurement, root mean square error (RMSE) between the observations and the predictions at the observed locations $$\on{RMSE}=\sqrt{\sum_{i=1}^n (y_i-\hat{y}_i)^2},$$ are calculated where the predictions $\hat{y}_i$ are taken to be the associated posterior mean. Since the SPDE approach aims to estimate the whole surface, smaller RMSE suggests the estimated surface is closer to the measurements at the observed locations. To measure the predictive performance, leave-one-out cross validation is employed using the embedded function within \texttt{R-inla}. The logarithmic score (Log Score) of prediction is defined as
$$\on{Log~Score}=-\frac{1}{N}\sum\limits_{i=1}^N \log[\pi(y_i|y_{-i})],$$
where $\pi(y_i|y_{-i})$ is the posterior predictive density of $y_i$ given all the other observations $y_{-i}$. Therefore the smaller Log Score is, the more certain we are with the predictions. Furthermore six meshes are built which are denoted by mesh $1$-$6$ respectively as shown in Figure \ref{cascadia_mesh}. The meshes are extended with coarse triangles to avoid boundary effect \citep{lindgren2011explicit}. The number of basis functions for each combination of mesh and SPDE method is shown in Table \ref{nb_study2}. Table \ref{cascadia_table_G_LS} presents the RMSEs and Log Scores using LFE-SPDE, BS-SPDE-G and BS-SPDE-LS with $d=2,3,4$ based on the six meshes respectively.

\begin{table}[htbp!]
\caption{Number of basis functions using LFE-SPDE and BS-SPDE-G/LS with $d=2,3,4$ based on six meshes}
\vspace{0.25cm}
\centering
\begin{tabular}{|c|c|c|c|c|}\hline
mesh & LFE-SPDE & BS-SPDE $d=2$ & BS-SPDE $d=3$ & BS-SPDE $d=4$\\ \hline
$1$ & $28$ & $89$ & $184$ & $313$ \\ \hline
$2$ & $70$ & $252$ & $547$ & $955$ \\ \hline
$3$ & $195$ & $749$ & $1663$ & $2937$ \\ \hline
$4$ & $244$ & $945$ & $2104$ & $3721$ \\ \hline
$5$ & $536$ & $2111$ & $4726$ & $8381$ \\ \hline
$6$ & $978$ & $3881$ & $8710$ & $15465$ \\ \hline
\end{tabular}
\label{nb_study2}
\end{table}

In general, it can be shown from Table \ref{cascadia_table_G_LS} that as the triangulation becomes denser, the estimations and predictions are more accurate using both LFE-SPDE and BS-SPDE-G/LS in most cases. For a particular mesh, the RMSEs and Log Scores of BS-SPDE-G/LS with $d\geq 2$ are generally smaller than those of LFE-SPDE. In terms of the number of basis functions, BS-SPDE-G/LS with $d\geq 2$ also demonstrate better performance than LFE-SPDE in most cases. For example, for region $4$, the RMSEs and Log Scores of BS-SPDE-G with $d=2$ based on mesh $4$ or BS-SPDE-G with $d=4$ based on mesh $2$ are much smaller than corresponding components in LFE-SPDE based on mesh $6$ while they have similar numbers of basis functions. BS-SPDE-LS approach shows similar properties with BS-SPDE-G in general. We notice that BS-SPDE-LS yields smaller RMSEs than BS-SPDE-G in most cases. However, in terms of Log Score, BS-SPDE-G performs better than BS-SPDE-LS in most cases for the other three regions except region $1$. Another point we want to mention is the sudden change in the model performance due to the refinement of mesh or increase of polynomial degree $d$. For example, the RMSE obtained using LFE-SPDE for region $2$ is about $15.87$ based on mesh $4$; it is decreased suddenly to only $0.037$ based on mesh $5$ or $0.012$ using BS-SPDE-G with $d=2$. We notice that the associated number of basis functions increases. This sudden change may because the finite elements or splines reach some level of degree of freedom that is enough to model the surface well.

\begin{figure}[htbp!]
\centering
\includegraphics[width=16cm]{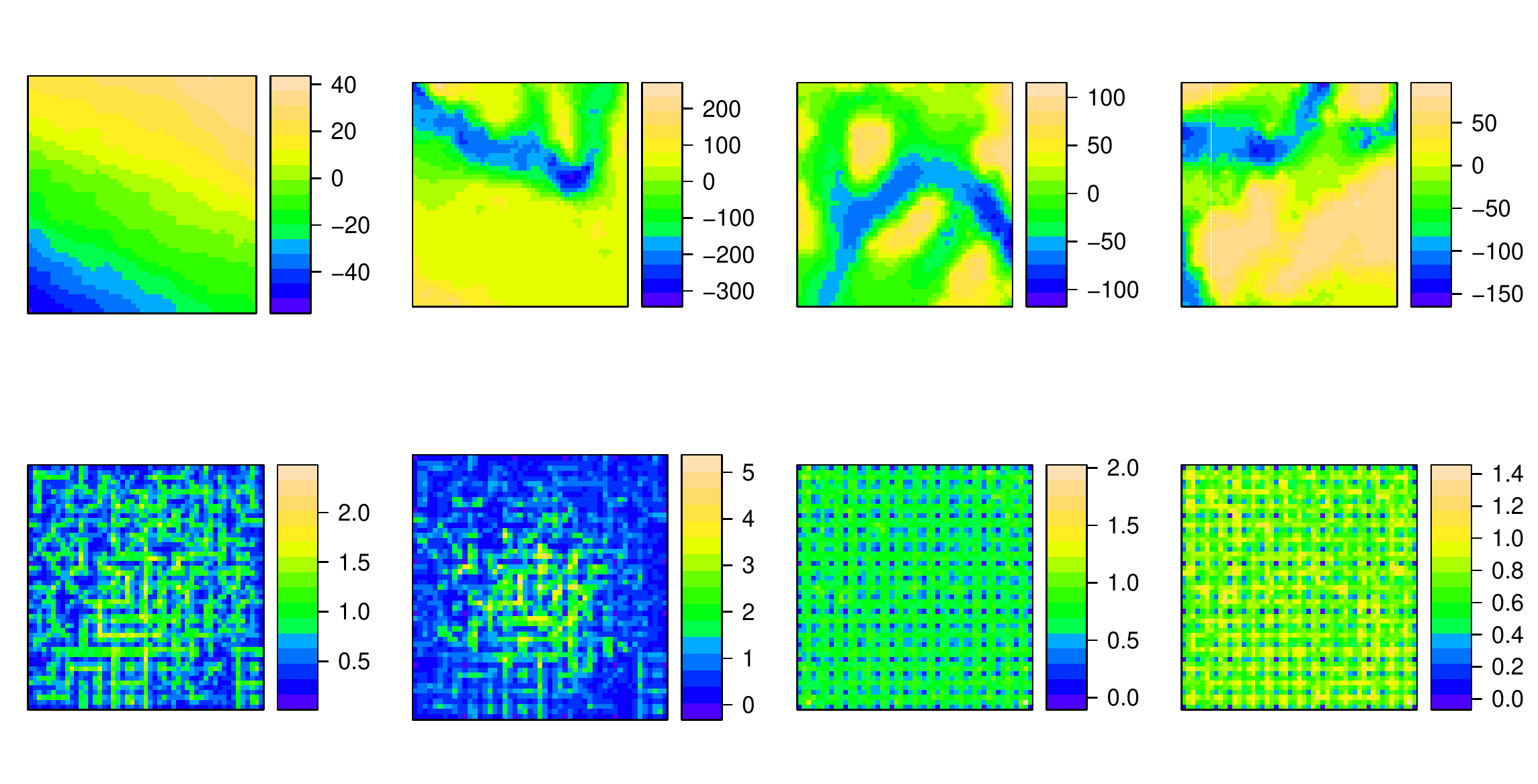}
\caption{Posterior mean (top) and standard deviation (bottom) for the regions $1$-$4$ (left to right), after selection of the appropriate approximation models (marked with asterisks in Table \ref{cascadia_table_G_LS})}
\label{4regions_pmsd_2by4}
\end{figure}

Based on Table \ref{cascadia_table_G_LS}, we can also select the models with good performance for continuous map reconstruction among the different combinations of meshes and SPDE approaches for each data set. Note that in most cases it is difficult to have a model with smallest RMSE and Log Score at the same time, so we only choose the one with relatively small RMSE and Log Score. In this way, the reconstructed map can be close to the elevations at the observed locations; meanwhile we are more confident with the predictions at the other locations. As marked with asterisks in Table \ref{cascadia_table_G_LS}, we choose BS-SPDE-LS with $d=4$ based on mesh $3$ for region $1$, BS-SPDE-LS with $d=3$ based on mesh $3$ for region $2$, BS-SPDE-G with $d=4$ based on mesh $3$ for region $3$, and BS-SPDE-G with $d=4$ based on mesh $3$ for region $4$. Note that for region $1$, BS-SPDE-LS with $d=4$ based on mesh $6$ yields both smallest RMSE and Log Score among all the combinations. However the associated computational cost is much heavier than the others. There is some trade off between model performance and computational cost. Hence we select the one with relatively good performance and is also computationally efficient. Then the posterior means and standard deviations of the four regions predicted using the respective selected models are displayed in Figure \ref{4regions_pmsd_2by4}. The posterior means in general capture the main features of the corresponding regions and the posterior standard deviations provide uncertainty estimates of the predictions. Note that the selection rule of predictive model here is quite simple and subjective. More appropriate model selection techniques can be employed in application.

\begin{table}[htbp!]
\caption{Study $2$: RMSE and Log Score (RMSE$|$Log Score) using LFE-SPDE, BS-SPDE-G and BS-SPDE-LS with $d=2,3,4$. The values in red colors are obtained based on similar number of basis functions between 945 to 978. The values marked with asterisks $(*)$ are the selected model fit for map reconstruction}
\vspace{0.25cm}
\centering
\setlength{\tabcolsep}{2pt}
\begin{tabular}{|c|c|c|c|c|c|c|c|}\hline 
\multirow{2}{*}{mesh} & \multirow{2}{*}{LFE-SPDE} & \multicolumn{3}{|c|}{BS-SPDE-G} & \multicolumn{3}{|c|}{BS-SPDE-LS} \\ \cline{3-8}
& & $d=2$ & $d=3$ & $d=4$ & $d=2$ & $d=3$ & $d=4$\\ \hline
\multicolumn{8}{|c|}{region $1$}\\ \hline 
$1$  & $1.85|2.07$ & $1.41|1.84$ & $1.18|1.73$ & $1.11|1.73$ & $1.41|1.84$ & $1.20|1.72$ & $1.16|1.71$ \\ \hline
$2$  & $1.17|1.71$ & $0.76|1.53$ & $0.55|1.47$ & \color{red}$0.29|1.21$ & $0.88|1.65$ & $0.81|1.70$ & \color{red}$0.74|1.73$ \\ \hline
$3$  & $0.83|2.01$ & $0.46|1.48$ & $0.33|1.42$ & $0.16|1.09$ & $0.43|1.64$ & $0.048|1.18$ & $0.027|0.98*$ \\ \hline
$4$  & $0.79|1.61$ & \color{red}$0.46|1.48$ & $0.34|1.43$ & $0.17|1.16$ & \color{red}$0.23|1.43$ & $0.039|1.08$ & $0.028|0.99$ \\ \hline
$5$  & $0.62|1.51$ & $0.49|1.50$ & $0.42|1.45$ & $0.36|1.38$ & $0.024|1.51$ & $0.022|1.30$ & $0.021|1.08$ \\ \hline
$6$  & \color{red}$0.63|1.52$ & $0.50|1.51$ & $0.49|1.48$ & $0.44|1.53$ & $0.021|1.74$ & $0.020|1.19$ & $0.019|0.92$ \\ \hline
\multicolumn{8}{|c|}{region $2$}\\ \hline 
$1$  & $79.81|5.82$ & $66.01|5.66$ & $52.26|5.41$ & $47.01|5.30$ & $66.41|5.65$ & $52.21|5.42$ & $45.81|5.27$ \\ \hline
$2$  & $36.20|5.02$ & $21.27|4.53$ & $9.09|3.86$ & \color{red}$0.02|0.50$ & $22.47|4.58$ & $14.52|4.23$ & \color{red}$12.48|4.10$ \\ \hline
$3$  & $17.24|4.34$ & $0.013|0.51$ & $0.0091|0.43$ & $0.0088|0.45$ & $0.015|0.43$ & $0.011|0.41*$ & $0.0094|0.50$ \\ \hline
$4$  & $15.87|4.28$ & \color{red}$0.012|0.52$ & $0.010|0.46$ & $0.010|0.50$ & \color{red}$0.013|0.47$ & $0.0098|0.46$ & $0.0091|0.54$ \\ \hline
$5$  & $0.037|1.85$ & $0.013|0.85$ & $0.010|0.55$ & $0.010|0.48$ & $0.0086|0.61$ & $0.0069|0.55$ & $0.0065|0.57$ \\ \hline
$6$  & \color{red}$0.032|1.98$ & $0.011|0.72$ & $0.010|0.55$ & $0.012|0.53$ & $0.0075|0.71$ & $0.0062|0.64$ & $0.0073|0.81$ \\ \hline
\multicolumn{8}{|c|}{region $3$}\\ \hline 
$1$  & $41.88|5.18$ & $37.52|5.09$ & $30.83|4.89$ & $25.44|4.69$ & $37.52|5.10$ & $30.51|4.89$ & $25.21|4.68$ \\ \hline
$2$  & $26.64|4.72$ & $11.62|3.93$ & $6.06|3.43$ & \color{red}$0.021|0.45$ & $11.70|3.94$ & $7.83|3.59$ & \color{red}$0.029|0.52$ \\ \hline
$3$  & $11.45|3.95$ & $0.028|0.74$ & $0.016|0.50$ & $0.015|0.48*$ & $0.022|0.58$ & $0.021|0.56$ & $0.017|0.61$ \\ \hline
$4$  & $10.18|3.86$ & \color{red}$0.025|0.72$ & $0.017|0.53$ & $0.015|0.48$ & \color{red}$0.021|0.55$ & $0.019|0.60$ & $0.016|0.59$ \\ \hline
$5$  & $7.21|3.66$ & $0.021|1.01$ & $0.018|0.64$ & $0.016|0.50$ & $0.014|0.69$ & $0.012|0.68$ & $0.012|0.72$ \\ \hline
$6$  & \color{red}$0.053|2.08$ & $0.019|0.99$ & $0.019|0.80$ & $0.019|0.64$ & $0.012|0.82$ & $0.011|0.79$ & $0.012|0.96$ \\ \hline
\multicolumn{8}{|c|}{region $4$}\\ \hline 
$1$  & $47.12|5.30$ & $35.98|5.07$ & $28.46|4.83$ & $22.95|4.61$ & $36.02|5.07$ & $28.46|4.83$ & $22.73|4.59$ \\ \hline
$2$  & $26.36|4.71$ & $12.56|4.02$ & $8.08|3.71$ & \color{red}$0.020|0.48$ & $12.47|4.01$ & $9.24|3.78$ & \color{red}$0.033|0.55$ \\ \hline
$3$  & $13.85|4.13$ & $0.023|0.62$ & $0.016|0.52$ & $0.013|0.43*$ & $0.019|0.55$ & $0.016|0.54$ & $0.014|0.56$ \\ \hline
$4$  & $12.42|4.04$ & \color{red}$0.021|0.76$ & $0.017|0.54$ & $0.014|0.46$ & \color{red}$0.019|0.54$ & $0.015|0.56$ & $0.014|0.59$ \\ \hline
$5$  & $0.061|2.19$ & $0.019|0.93$ & $0.017|0.64$ & $0.015|0.49$ & $0.012|0.63$ & $0.011|0.63$ & $0.010|0.68$ \\ \hline
$6$  & \color{red}$0.049|2.14$ & $0.017|0.93$ & $0.018|0.78$ & $0.017|0.58$ & $0.011|0.77$ & $0.010|0.77$ & $0.010|0.88$ \\ \hline
\end{tabular}
\label{cascadia_table_G_LS}
\end{table}

\subsection{Spatial analysis of ozone levels data over Eastern USA}
\label{subsec:ozone}

\begin{figure}[htbp!]
\centering
\includegraphics[width=10cm]{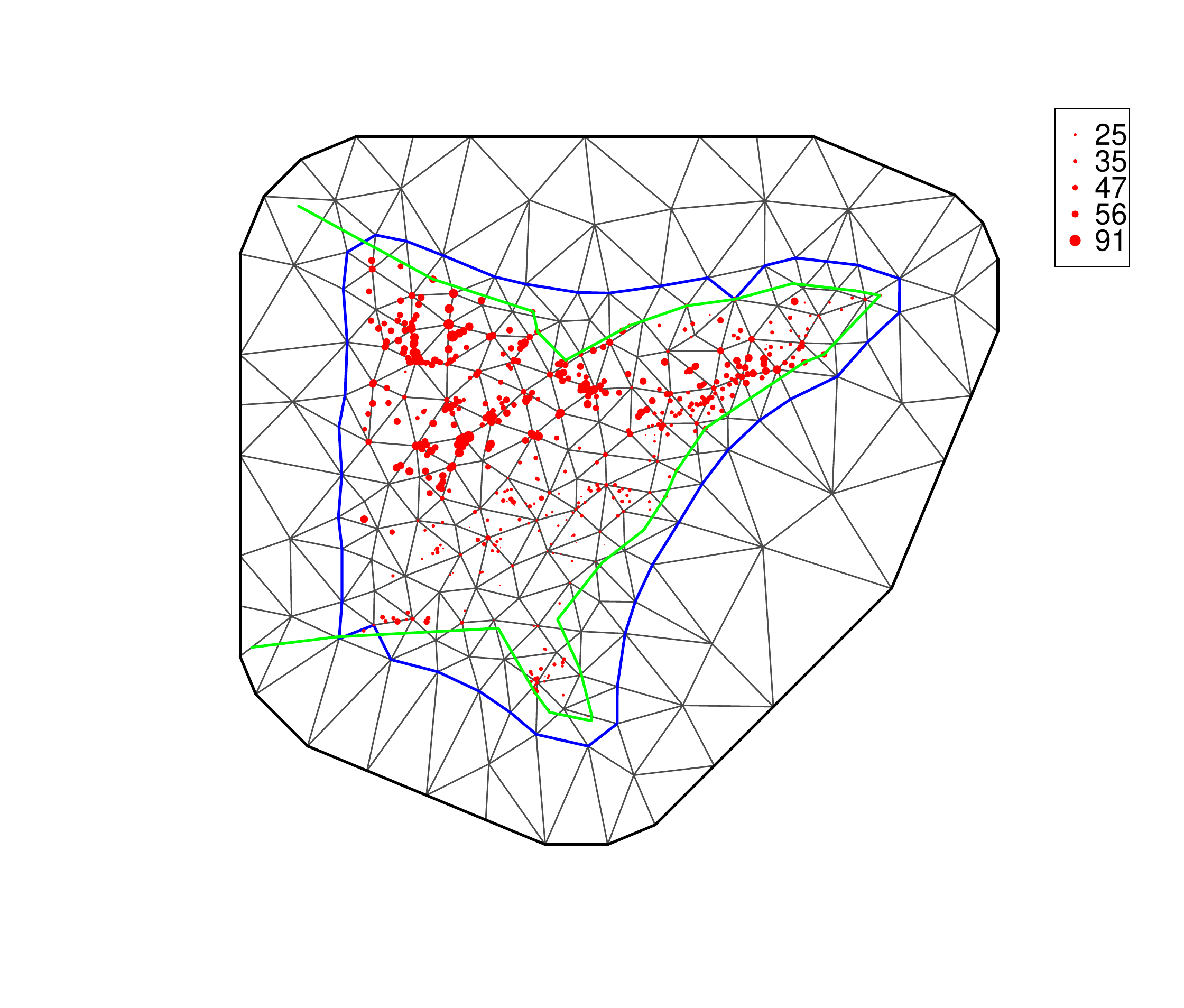}
\caption{Triangulation over Eastern United States; green line: U.S. boundary; red dots: locations of ozone monitoring stations; size proportional to the ozone levels in ppb (parts per billion)}
\label{oze_mesh}
\end{figure}

In this section, we analyse a data set of ozone levels at a certain hour in one of days in September, 2005 around the Eastern United States, which is available from the Air Explorer Database of Environmental Protection Agency (EPA), using the non-stationary BS-SPDE-G method. 
The data set has $546$ locations where ozone levels are recorded. As shown in Figure $\ref{oze_mesh}$, the observations of ozone concentration are distributed unevenly and the domain is irregular. Denote the ozone levels by $z_i$ and the associated locations by $\mathbf{s}_i=(x_i,y_i)$, $i=1,...,546$. We consider a simple spatial model
$$z_i \sim b_0 + f(\mathbf{s}_i),~~ i=1,...,546,$$
where $b_0$ is the intercept and the spatial effect $f(\mathbf{s}_i)$ is assumed to be a non-stationary GF generated by the non-stationary version SPDE $(\ref{spde_ns})$, approximated with bivariate splines in $S_d^0(\boldsymbol{\Delta})$ with $d=1,2,3,4,5$. The triangulation $\boldsymbol{\Delta}$ is shown in Figure $\ref{oze_mesh}$.

The non-stationary parameters $\tau(\mathbf{u})$ and $\kappa(\mathbf{u})$ are represented with two-dimensional B-splines that have $n_x$ and $n_y$ basis functions in the $x$-direction and $y$-direction respectively. Therefore at any location $\mathbf{s}=(x,y)$, the basis functions of the associated B-spline can be calculated as $B_{lk}(\mathbf{s})=B_l^{x}(x)B_k^{y}(y)$, where $B_{l}^x(\cdot)$ and $B_k^y(\cdot)$ are the basis functions in $x$ and $y$ directions respectively, for $l=1,...,n_x$ and $k=1,...,n_y$. Hence there are $n_xn_y$ basis functions in total for each of the parameters  $\kappa^2(\cdot)$ and $\tau(\cdot)$.
We consider $12$ models $A$-$L$ with different combinations of the number of basis functions in Table $\ref{oze_nb}$. Note that with one basis function, the B-spline is constant so that model $A$ corresponds to the stationary SPDE model $(\ref{spde})$, and the number of basis functions represents the number of basis functions in both $x$-direction and $y$-direction; for example in model $C$, there are $3$ basis functions for $\tau(\cdot)$ in $x$-direction as well as $y$-direction so that there are actually $3\times3=9$ basis functions for $\tau(\cdot)$.

\begin{table}[htbp!]
\caption{Number of basis functions for the B-spline in each direction for the parameters $\kappa^2(\cdot)$ and $\tau(\cdot)$}
\vspace{0.25cm}
\centering
\begin{tabular}{c|cccccccccccc}
\hline
 & $A$ & $B$ & $C$ & $D$ & $E$ & $F$ & $G$ & $H$ & $I$ & $J$ & $K$ & $L$ \\\hline
$\kappa^2(\cdot)$ & $1$ & $1$ & $1$ & $1$ & $1$ & $2$ & $3$ & $4$ & $5$ & $2$ & $3$ & $4$ \\\hline
$\tau(\cdot)$ & $1$ & $2$ & $3$ & $4$ & $5$ & $1$ & $1$ & $1$ & $1$ & $2$ & $3$ & $4$\\
\hline
\end{tabular}
\label{oze_nb}
\end{table}

\begin{figure}[htbp!]
\centering
\includegraphics[width=15cm]{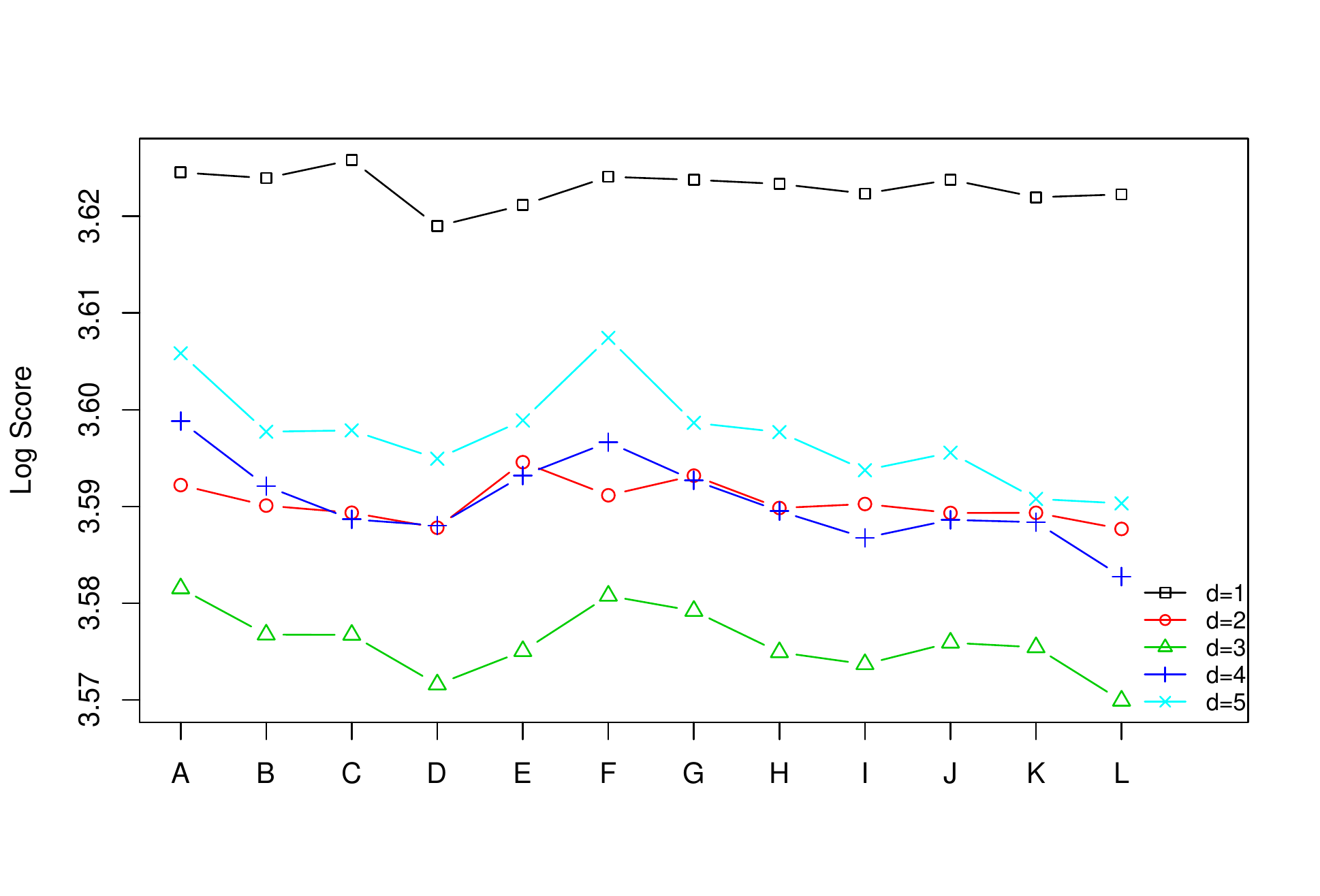}
\caption{Log Scores for the models $A$-$L$ using BS-SPDE-G with $d=1,...,5$}
\label{oze_logscore}
\end{figure}

To measure the fit and predictive performance and select the appropriate representations for $\kappa^2(\cdot)$ and $\tau(\cdot)$, we employ the leave-one-out cross validation and aim to find the model with the smallest Log Score.
Figure $\ref{oze_logscore}$ presents the Log Scores of the $12$ models for the two parameters $\kappa^2(\cdot)$ and $\tau(\cdot)$ as shown in Table $\ref{oze_nb}$ using BS-SPDE-G approach with $d=1,2,3,4,5$ respectively. It is easy to see that the Log Scores obtained from BS-SPDE-G with higher $d$ are generally smaller than those obtained from BS-SPDE-G with lower $d$. Using BS-SPDE-G with a specific $d$, the Log Scores for different representations of $\kappa^2(\cdot)$ and $\tau(\cdot)$ are different. In general the non-stationary models $B$-$L$ yield smaller Log Scores than the stationary model $A$ and the overall smallest Log Score is obtained with model $L$ and BS-SPDE-G $d=3$. Hence model $L$ is chosen to be the model for the non-stationarity in this case for further prediction. As shown in Table $\ref{oze_nb}$, model $L$ corresponds to $4$ basis functions for $\kappa^2(\cdot)$ and $4$ basis functions for $\tau(\cdot)$ in both $x$ and $y$ direction. This suggests that both $\kappa^2(\cdot)$ and $\tau(\cdot)$ display spatial variation over the domain. Note that the number of parameters are not taken into account for model selection here. In fact, we notice that the Log Score obtained using BS-SPDE-G $d=3$ with model $D$ is only slightly higher than model $L$ while the number of parameters used to represent the non-stationarity is only half of model $L$. A proper model selection technique would account for it, e.g. AIC and BIC, but this is beyond the scope of this paper.

\begin{figure}[htbp!]
\centering
\includegraphics[width=15cm]{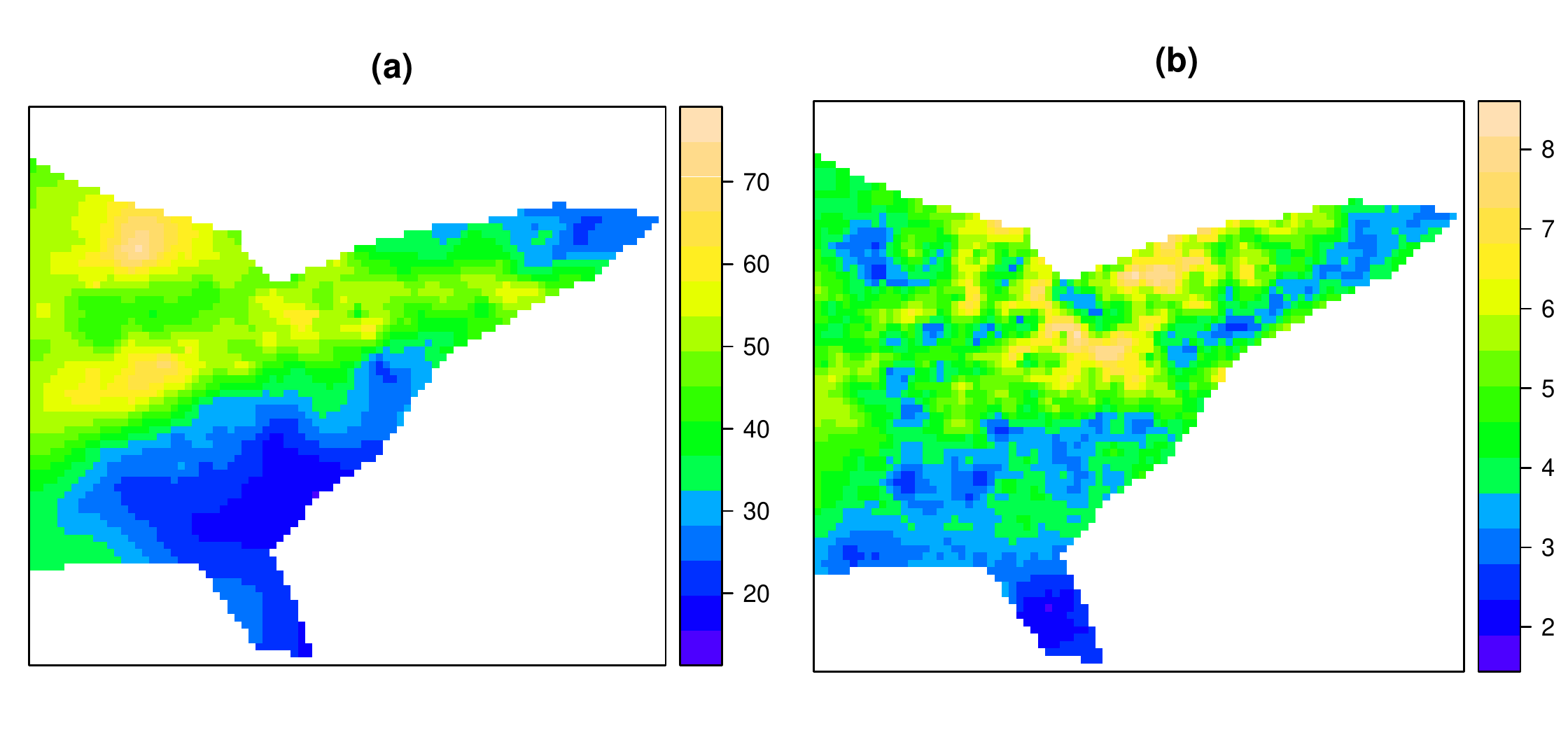}
\caption{(a) Posterior mean; (b) posterior standard deviation: ozone levels over Eastern United States predicted using BS-SPDE-G with $d=3$ and model $L$ for the two parameters $\kappa^2(\cdot)$ and $\tau(\cdot)$}
\label{oze_pmsd}
\end{figure}

Then we apply the non-stationary model $L$ for the two parameters $\kappa^2$ and $\tau$ and predict ozone levels over the Eastern United States using the BS-SPDE-G approach with $d=3$. Figure $\ref{oze_pmsd}$ displays the posterior mean and standard deviation of the predictions given the observations presented in Figure $\ref{oze_mesh}$. As we can see, the predicted ozone level is low in the south-east corner and at the top of the north-east corner and high in the north and middle area, which is consistent with the observations. Furthermore, the posterior predictive standard deviation shows some spatial variation over the entire domain because of the irregular distribution of the observations and the non-stationarity of the SPDE model.

\section{Discussion}
\label{sec:discuss}
We have shown that polynomial basis of order greater than one can be easily implemented in the SPDE framework for GFs using bivariate splines. Both the theoretical results and numerical simulations have demonstrated that this new approach has its advantages over the linear finite element approach in many applications in terms of both approximation accuracy and computational efficiency.  By using higher degree representations, we can also implement the least squares solutions to the SPDE $(\ref{spde})$ for $\alpha=2$. This should be more computationally efficient to make inference than the corresponding Galerkin solutions due to the different sparsity structures of the two solutions. In terms of application, we have shown that the SPDE approach can be applied to the spatial modelling of bathymetry. However, the current commonly used mapping tools, e.g. Generic Mapping Tools (GMT) used by NOAA \citep{Eakins10}, do not include uncertainty estimates of the maps. GMT also requires high smoothness conditions, see \cite{smith1990gridding} and \cite{wessel1998interpolation}, that may not be appropriate for the bathymetry/topography. Hence the computationally efficient SPDE approach has promising potential in spatial mapping.

There is still room for further investigation in the application of bivariate splines in the SPDE approach. It has been suggested by the numerical simulations that the degree of polynomial basis has an impact on the performance of the SPDE approach so that choosing appropriate degrees is essential to the SPDE approach to capture global and local spatial characteristics. In fact, within the framework of bivariate splines, the degree of polynomials can be different over different triangles. \cite{hu2007bivariate} proposed a new spline method which allows automatic degree raising over triangles of interest. This new method is able to solve linear PDEs very effectively and efficiently. Another extension to manifolds could be considered. \cite{lai2009triangulated} discussed the application of spherical splines in geopotential approximation where the techniques of triangulated spherical splines can be applied to represent the Mat\'ern fields on manifolds. Furthermore, when $\alpha$ is larger than $2$, which means the smoothness parameter $\nu$ in $(\ref{matern})$ increases as well, smoother sample paths of the Mat\'ern fields are expected according to \cite{paciorek2004nonstationary}. However, it is quite difficult to implement higher orders of smoothness in conventional finite element representations. But within the bivariate splines framework, higher orders of smoothness conditions can be implemented easily by imposing linear constraints on the B-coefficients \citep{lai2007spline} which leads to another potential extension of this approach allowing smoother representations of the GFs. However the locally supported bases for smoother bivariate spaces $S_d^r(\boldsymbol{\Delta})$ for $r>0$ are difficult to construct and the Bayesian inference with large number of linear constraints are still open to research. Hence this needs to be investigated further in order to implement the SPDE approach with bivariate splines of higher smoothness.

\begin{appendix}
\section{Appendix}
\label{app:appendix}
We include some details about bivariate splines relevant to this paper. Then the sketch of proofs for Theorem $\ref{Q_theorem}$ and Propositions $\ref{convergence_rate}$ - $\ref{Markov_approx}$ are provided.
\subsection{Preliminaries on bivariate splines}
\label{app:bs}
To evaluate a polynomial of degree $d$ in B-form over any triangle, say $p=\sum_{i+j+k=d}c_{ijk}B_{ijk}^d$, at the point $\mathbf{v}=(x,y)$ whose barycentric coordinates are $b=(b_1,b_2,b_3)$ with $b_1+b_2+b_3=1$, let $c_{ijk}^{(0)}=c_{ijk}$ and for all $l=1,...,d$,
$$c_{ijk}^{(l)}=b_1c_{i+1,j,k}^{(l-1)}+b_2c_{i,j+1,k}^{(l-1)}+b_3c_{i,j,k+1}^{(l-1)}.$$
For $i+j+k=d-l$, we have
$$p(\mathbf{v})=\sum_{i+j+k=d-l}c_{ijk}^{(l)}B_{ijk}^{d-l}(\mathbf{v}),$$ for all $0\leq l \leq d$. In particular,
$p(\mathbf{v})=c_{000}^{(d)}.$ This is called the de Casteljau algorithm \citep{lai2007spline}.

Each vector $\mathbf{u}$ can be uniquely described by a triple $(a_1,a_2,a_3)$ called directional coordinates of $\mathbf{u}$, that is $a_i=\alpha_i-\beta_i$, $i=1,2,3$, where $(\alpha_1,\alpha_2,\alpha_3)$ and $(\beta_1,\beta_2,\beta_3)$ are the barycentric coordinates of two points $\boldsymbol{\omega}$ and $\tilde{\boldsymbol{\omega}}$ such that $\mathbf{u} = \boldsymbol{\omega}-\tilde{\boldsymbol{\omega}}$. It is easy to see that the barycentric coordinates of a point sum to 1, while the directional coordinates of a vector sum to 0. Suppose $\mathbf{u}$ is a vector in $\mathbb{R}^2$ whose directional coordinates are $a=(a_1,a_2,a_3)$, then for $i+j+k=d$, we define the directional derivative of $B_{ijk}^d$ at location $\mathbf{v}$ with respect to directional vector $\mathbf{u}$ to be
\begin{equation}\label{dB}
D_uB_{ijk}^d(\mathbf{v})=d\left[a_1B_{i-1,j,k}^{d-1}(\mathbf{v})+a_2B_{i,j-1,k}^{d-1}(\mathbf{v})+a_3B_{i,j,k-1}^{d-1}(\mathbf{v})\right].
\end{equation}

The integrals and inner products of the Bernstein polynomials can be calculated precisely as presented in the following lemma.
\begin{lemma}\label{integral}
Let $p = \sum\limits_{i+j+k=d} c_{ijk}B_{ijk}^d$ be a polynomial of degree $d$ on triangle $T$ (with area $A_T$), then
\begin{equation}\label{intB}
\int_T p(x,y)dxdy=\frac{A_T}{\tbinom{d+2}{2}}\sum_{i+j+k=d}c_{ijk}.
\end{equation}
Let $q = \sum\limits_{\nu+\mu+\kappa=d} \tilde{c}_{\nu\mu\kappa}B_{\nu\mu\kappa}^d$ be another polynomial of degree $d$ on triangle $T$, then the inner product of $p$ and $q$ is
\begin{equation}\label{intBB}
\int_T p(x,y) q(x,y)dxdy=\frac{A_T}{\tbinom{2d}{d}\tbinom{2d+2}{2}}\sum_{\substack{i+j+k=d \\ \nu+\mu+\kappa=d}}\tbinom{i+\nu}{i}\tbinom{j+\mu}{j}\tbinom{k+\kappa}{k}c_{ijk}\tilde{c}_{\nu\mu\kappa}.
\end{equation}
\end{lemma}

For the spline space $S_d^0(\boldsymbol{\Delta})$, the domain points are defined to be the set
$$\mathcal{D}_{d,\boldsymbol{\Delta}}=\{\xi_{ijk}=(i\mathbf{v}_1+j\mathbf{v}_2+k\mathbf{v}_3)/d, i+j+k=d,T=\langle \mathbf{v}_1,\mathbf{v}_2,\mathbf{v}_3\rangle \in \boldsymbol{\Delta}\}.$$
Therefore the spline function can also be denoted by
$$s|_T=\sum_{\xi\in\mathcal{D}_{d,T}}c_{\xi}B_{\xi}^{T,d},$$
where $B_{\xi}^{T,d}$ stands for $B_{ijk}^{T,d}$ for $\xi=\xi_{ijk}\in\mathcal{D}_{d,T}$ and $c_{\xi}$ is the corresponding B-coefficient $c_{ijk}$. Note that since $s$ is continuous, if $\xi$ lies on an edge shared by two different triangles $T$ and $\tilde{T}$, then the corresponding coefficients $c_{\xi}$ for $s|_T$ and $s|_{\tilde{T}}$ should be the same.
Then we show that the basis for $S_d^0(\boldsymbol{\Delta})$ can be constructed easily with spline functions in $S_d^0(\boldsymbol{\Delta})$ with specific B-coefficients. For each $\xi\in\mathcal{D}_{d,\boldsymbol{\Delta}}$, let $\psi_{\xi}$ be the spline in $S_d^0(\boldsymbol{\Delta})$ having all zero B-coefficients except for $c_\xi=1$, then we have the following result
\begin{lemma}\label{Sd0basis}
The set of splines $\mathcal{B}=\{\psi_{\xi},\xi\in\mathcal{D}_{d,\boldsymbol{\Delta}}\}$ forms a basis for the spline space $S_d^0(\boldsymbol{\Delta})$ which satisfies
$\psi_{\xi}(\mathbf{v})\geq 0$ and
$\sum_{\xi\in\mathcal{D}_{d,\boldsymbol{\Delta}}}\psi_{\xi}(\mathbf{v})=1$
for all $\mathbf{v}\in\Omega$.
\end{lemma}
It is obvious that $\psi_{\xi}$ is identically zero on all triangles that do not contain $\xi$ since the corresponding B-coefficients are all zeros so that $\psi_{\xi}$ is locally supported.

\subsection{Proof of Theorem $\ref{Q_theorem}$}
\label{app:th1}
The derivation of $\mathbf{Q}_1$, $\mathbf{Q}_2^{G}$ and $\mathbf{Q}_{\alpha}$ for $\alpha\geq 3$ can be found directly from Appendix D.3.1 of \cite{lindgren2011explicit}. Here we only present briefly how to calculate the least squares solution $\mathbf{Q}_2^{LS}$ and the matrix components $\mathbf{M}$, $\mathbf{K}$ and $\mathbf{R}$.

When $\alpha=2$, by plugging the bivariate spline representation of $x(\mathbf{u})$ in to the equality $(\ref{spde_solver})$, we have
$$\langle\phi, \sum\limits_{h=1}^m(\kappa^2-\Delta)\tau\psi_h w_h\rangle \deq \langle \phi,W\rangle,$$
for any appropriate test functions $\phi$. By choosing a set of test functions to be $\phi_h=(\kappa^2-\Delta)\psi_h$, we have
\begin{equation}\label{test_linear_eqs}
\langle(\kappa^2-\Delta)\psi_t, \sum\limits_{s=1}^m(\kappa^2-\Delta)\tau\psi_sw_s\rangle \deq \langle (\kappa^2-\Delta)\psi_t,W\rangle,~~~t=1,...,m.
\end{equation}
The left hand side is
$$\sum\limits_{s=1}^m\tau(\kappa^4\langle\psi_t,\psi_s\rangle+2\kappa^2\langle\nabla\psi_t,\nabla\psi_s\rangle+\langle\Delta\psi_s,\Delta\psi_t\rangle)w_s,$$
by applying the stochastic Green's first identity along with the Neumann boundary conditions. The integral on the right hand side is in fact Gaussian with mean zero and covariance matrix whose $(t,s)$-th element is
$$\on{Cov}(\langle (\kappa^2-\Delta)\psi_t,W\rangle,\langle (\kappa^2-\Delta)\psi_s,W\rangle)=\kappa^4\langle\psi_t,\psi_s\rangle+2\kappa^2\langle\nabla\psi_t,\nabla\psi_s\rangle+\langle\Delta\psi_s,\Delta\psi_t\rangle.$$
Then we can write $(\ref{test_linear_eqs})$ in the matrix form as
$$\tau(\kappa^4\mathbf{M}+2\kappa^2\mathbf{K}+\mathbf{R})\mathbf{w}\sim N(\mathbf{0},\kappa^4\mathbf{M}+2\kappa^2\mathbf{K}+\mathbf{R}),$$
where the $(t,s)$-th entry of the matrices $\mathbf{M}$, $\mathbf{K}$ and $\mathbf{R}$ are respectively $\mathbf{M}_{ts}=\langle\psi_t,\psi_s\rangle$, $\mathbf{K}_{ts}=\langle\nabla\psi_t,\nabla\psi_s\rangle$, and
$\mathbf{R}_{ts}=\langle\Delta\psi_t,\Delta\psi_s\rangle$. which are usually called mass matrix, stiffness matrix and roughness matrix in bivariate spline literature. Therefore it is easy to show that the precision matrix of $\mathbf{w}$ is $\mathbf{Q}=\tau^2(\kappa^4\mathbf{M}+2\kappa^2\mathbf{K}+\mathbf{R})$.

Following the Lemma $\ref{integral}$ and $\nabla p=\sum_{i+j+k=d}c_{ijk}\nabla B_{ijk}^d$ , $\Delta p=\sum_{i+j+k=d}c_{ijk}\Delta B_{ijk}^d$ for any $p=\sum_{i+j+k=d}c_{ijk}B_{ijk}^d$, we have the contribution of triangle $T$ to the $(t,s)$-th entry of $\mathbf{M}$, $\mathbf{K}$ and $\mathbf{R}$ for $t,s=1,...,m$ are
$$\mathbf{M}_{ts}|_{T}=\langle \psi_t,\psi_s\rangle_{T}=\mathbf{c}'_t|_{T}M_T\mathbf{c}_s|_T,$$
$$\mathbf{K}_{ts}|_{T}=\langle \nabla\psi_t,\nabla\psi_s\rangle_{T}=\mathbf{c}'_t|_{T}K_T\mathbf{c}_s|_{T},$$
$$\mathbf{R}_{ts}|_{T}=\langle \nabla\psi_t,\nabla\psi_s\rangle_{T}=\mathbf{c}'_t|_{T}R_T\mathbf{c}_s|_{T},$$
where $M_T$, $K_T$ and $R_T$ are defined in Theorem $\ref{Q_theorem}$, and $\mathbf{c}_h|_T$ is the column vector of B-coefficients of $\psi_h$ associated with triangle $T$, $h=1,...,m$. Then it is followed that
$$\mathbf{M}_{ts}=\sum_T\mathbf{M}_{ts}|_{T}=\mathbf{c}'_t\mathbf{M_0}\mathbf{c}_s,~~~\mathbf{K}_{ts}=\sum_T\mathbf{K}_{ts}|_{T}=\mathbf{c}'_t\mathbf{K_0}\mathbf{c}_s,~~~\mathbf{R}_{ts}=\sum_T\mathbf{M}_{ts}|_{T}=\mathbf{c}'_t\mathbf{R_0}\mathbf{c}_s,$$
where $\mathbf{M_0}=\on{diag}(\mathbf{M}_T,T\in\boldsymbol{\Delta})$, $\mathbf{K_0}=\on{diag}(\mathbf{K}_T,T\in\boldsymbol{\Delta})$ and $\mathbf{R_0}=\on{diag}(\mathbf{R}_T,T\in\boldsymbol{\Delta})$. Therefore we have the following simple matrix representation that
$$\mathbf{M}=\mathbf{C}'\mathbf{M_0}\mathbf{C},~~~ \mathbf{K}=\mathbf{C}'\mathbf{K_0}\mathbf{C}, ~~~
\mathbf{R}=\mathbf{C}'\mathbf{R_0}\mathbf{C}.$$

\subsection{Proof of Proposition $\ref{convergence_rate}$}
\label{app:prop1}
For $1\leq q \leq \infty$ and $d\geq 1$, the associated Sobolev space on $\Omega$ in $\mathbb{R}^2$ is defined by
$$W_q^{d}(\Omega)=\{f: \| f \|_{d,q,\Omega}<\infty\},$$
where
$$\| f\|_{d,q,\Omega}=\begin{cases} \left(\sum\limits_{k=0}^{d}|f|_{k,q,\Omega}^q\right)^{1/q}, & 1\leq q < \infty\\ \sum\limits_{k=0}^{d}|f|_{k,\infty,\Omega}, & q=\infty, \end{cases}$$ with
$$|f|_{k,q,\Omega}=\begin{cases} \left(\sum\limits_{\nu+\mu=k}\| D_x^{\nu}D_y^{\mu}f\|_{q,\Omega}^q\right)^{1/q}, & 1\leq q < \infty\\ \max\limits_{\nu+\mu=k}\| D_x^{\nu}D_y^{\mu}f\|_{\infty,\Omega}, & q=\infty, \end{cases}$$
and
$$\| f \|_{q,\Omega}=\begin{cases} \left(\int_{\Omega}|f(u)du|^q\right)^{1/q}, & 1\leq q < \infty,\\ \esssup_{u\in\Omega} |f(u)|, & q=\infty.\end{cases}$$

Let $f_{\boldsymbol{\Delta}}(\mathbf{s})$ be the $H^1$-orthogonal projection of $f\in H^1\cap W_2^{m+1}(\Omega)$ onto the bivariate spline space $S_d^0(\boldsymbol{\Delta})$, it follows that
\begin{eqnarray*}
\begin{aligned}
\int_\Omega f(\mathbf{s})Lx_{\boldsymbol{\Delta}}(\mathbf{s})d\mathbf{s} & = \int_\Omega (f(\mathbf{s})-f_{\boldsymbol{\Delta}}(\mathbf{s}))Lx_{\boldsymbol{\Delta}}(\mathbf{s})d\mathbf{s} + \int_\Omega f_{\boldsymbol{\Delta}}(\mathbf{s})Lx_{\boldsymbol{\Delta}}(\mathbf{s})d\mathbf{s}\\
& = \int_\Omega f_{\boldsymbol{\Delta}}(\mathbf{s})Lx_{\boldsymbol{\Delta}}(\mathbf{s})d\mathbf{s}\\
& = \int_\Omega f_{\boldsymbol{\Delta}}(\mathbf{s})dW(\mathbf{s}),
\end{aligned}
\end{eqnarray*}
where the second equality follows from the orthogonality of $f(\mathbf{s})-f_{\boldsymbol{\Delta}}(\mathbf{s})$ to $S_d^0(\boldsymbol{\Delta})$ with respect to $H^1$ inner product. Then we have
$$\int_\Omega f(\mathbf{s})L(x(\mathbf{s})-x_{\boldsymbol{\Delta}}(\mathbf{s}))d\mathbf{s}=\int_\Omega (f(\mathbf{s})-f_{\boldsymbol{\Delta}}(\mathbf{s}))dW(\mathbf{s}).$$
Hence it follows from the white noise integrals that
\begin{eqnarray*}
\begin{aligned}
\on{E}\left(\int_\Omega f(\mathbf{s})L(x(\mathbf{s})-x_{\boldsymbol{\Delta}}(\mathbf{s}))d\mathbf{s}\right)^2 & = \on{E}\left(\int_\Omega (f(\mathbf{s})-f_{\boldsymbol{\Delta}}(\mathbf{s}))dW(\mathbf{s})\right)^2\\
& = \int_\Omega(f(\mathbf{s})-f_{\boldsymbol{\Delta}}(\mathbf{s}))^2 d\mathbf{s}.
\end{aligned}
\end{eqnarray*}
Then it follows from standard results in bivariate splines literatures, for example Th. $5.19$ in \cite{lai2007spline} that under some suitable assumptions on the triangulation, we have for $1\leq m \leq d$,
$$\| f-f_{\boldsymbol{\Delta}} \|_{2,\Omega}\leq K|\boldsymbol{\Delta}|^{m+1}|f|_{m+1,2,\Omega}.$$

\subsection{Proof of Proposition $\ref{Markov_approx}$}
\label{app:prop2}
First of all, it is easy to see that
\begin{equation}
\label{FirstEq}
w_f' {\bf M} w_g=\langle f_{\boldsymbol{\Delta}},
g_{\boldsymbol{\Delta}}\rangle_{\boldsymbol{\Delta}}
=\sum_{T\in \boldsymbol{\Delta}}\int_T
f_{\boldsymbol{\Delta}}g_{\boldsymbol{\Delta}}dxdy
\end{equation}
since $f_{\boldsymbol{\Delta}},
g_{\boldsymbol{\Delta}} \in S^0_d(\boldsymbol{\Delta})$.
Next we can see
\begin{align*}
\mathbf{w}_f'\tilde{\mathbf{M}}\mathbf{w}_g
&= \sum_{T\in \boldsymbol{\Delta}}\sum_{\xi\in
\mathcal{D}_{d,T}} c_\xi( f_{\boldsymbol{\Delta}}) \sum_{\eta }
\int_T \phi_\xi \phi_\eta dxdy c_\xi(g_{\boldsymbol{\Delta}})\cr
&= \sum_{T\in \boldsymbol{\Delta}}\sum_{\xi\in
\mathcal{D}_{d,T}} \frac{A_T}{\tbinom{d+2}{2}}
 c_\xi( f_{\boldsymbol{\Delta}})
 c_\xi(g_{\boldsymbol{\Delta}}),
\end{align*}
where $\mathcal{D}_{d,T}=\{(i\mathbf{v}_1+j\mathbf{v}_2+k\mathbf{v}_3)/d,
i+j+k=d\}$ is the  set of associated domain points of triangle
$T=\langle \mathbf{v}_1,\mathbf{v}_2,\mathbf{v}_3\rangle$, $A_T$ is the
area of triangle $T$ and $c_\xi(s)$ is the B-coefficient of $s$.
When $f_{\boldsymbol{\Delta}}=C$ is a constant $C$, it is easy to see
that
$$
\sum_{\xi\in
\mathcal{D}_{d,T}} \frac{A_T}{\tbinom{d+2}{2}}
 c_\xi( f_{\boldsymbol{\Delta}})
 c_\xi(g_{\boldsymbol{\Delta}}) = \int_T f_{\boldsymbol{\Delta}}
 g_{\boldsymbol{\Delta}}dxdy
$$
and hence, we have
$$
\mathbf{w}_f'\tilde{\mathbf{M}}\mathbf{w}_g
=\sum_{T\in \boldsymbol{\Delta}}\int_T f_{\boldsymbol{\Delta}}
 g_{\boldsymbol{\Delta}}dxdy
$$
which is  $w_f' {\bf M} w_g$ by (\ref{FirstEq}). Similar when
$g_{\boldsymbol{\Delta}}$ is a piecewise constant. Also, when $d=1$,
this result follows from Lemma 1 in \cite{chen1985lumped}. We now
prove it for general $d\ge 1$.

We first note that
\begin{align*}
 c_\xi( f_{\boldsymbol{\Delta}} )  c_\xi( g_{\boldsymbol{\Delta}} )
=&
 ( c_\xi ( f_{\boldsymbol{\Delta}})- f_{\boldsymbol{\Delta}}(\xi ) )
 c_\xi(g_{\boldsymbol{\Delta}}) +
 f_{\boldsymbol{\Delta}}(\xi ) (c_\xi(g_{\boldsymbol{\Delta}})
 -g_{\boldsymbol{\Delta}}(\xi)) \cr
&+ f_{\boldsymbol{\Delta}}(\xi )g_{\boldsymbol{\Delta}}(\xi ).
\end{align*}
Then we claim that
$$
\sum_{\xi\in
\mathcal{D}_{d,T}} \frac{A_T}{\tbinom{d+2}{2}}
f_{\boldsymbol{\Delta}}(\xi )g_{\boldsymbol{\Delta}}(\xi )
\hbox{ approximates } \int_T f_{\boldsymbol{\Delta}}(x,y)g_{\boldsymbol{\Delta}}(x,y)dxdy.
$$
Indeed, let us recall the Bernstein-B\'ezier approximation of arbitrary
continuous function $F$ on $T$. That is, using Th. 2.45 in
\cite{lai2007spline}, we have
\begin{equation}
\label{Lai2.45}
\|F- B_d(F)\|_{T,\infty}\le \frac{|T|^2}{d}|F|_{2,T}
\end{equation}
where $B_d(F)= \sum_{\xi\in \mathcal{D}_{d,T}} F(\xi) B_\xi$ and
$B_\xi$ are the Bernestein-B\'ezier polynomials of degree $d$. Letting
$F(x,y)=f_{\boldsymbol{\Delta}}(x,y)g_{\boldsymbol{\Delta}}(x,y)$, we
have
\begin{align*}
&|\int_T F(x,y)dxdy- \int_T B_d(F)dxdy| \cr
=& |\int_T f_{\boldsymbol{\Delta}}(x,y)g_{\boldsymbol{\Delta}}(x,y)dxdy
- \sum_{\xi\in \mathcal{D}_{d,T}}
f_{\boldsymbol{\Delta}}(\xi )g_{\boldsymbol{\Delta}}(\xi )
\frac{A_T}{\tbinom{d+2}{2}} | \cr
\le & \frac{|T|^2}{d}
\int_T |f_{\boldsymbol{\Delta}}g_{\boldsymbol{\Delta}}|_{2,T}dxdy\cr
\le & K \frac{|T|^2}{d}
|f_{\boldsymbol{\Delta}}g_{\boldsymbol{\Delta}}|_{2,1,T}\cr
\le & K\frac{|T|^2}{d} |f_{\boldsymbol{\Delta}}|_{2,2,T}
|g_{\boldsymbol{\Delta}}|_{2,2,T},
\end{align*}
where we have used the fact that
 $f_{\boldsymbol{\Delta}}g_{\boldsymbol{\Delta}}$ is a polynomial of
degree $2d$ in the second inequality and
the Cauchy-Schwarz inequality in the last
inequality. This finishes the proof of the claim.

Next we consider
$$
I_1(T):=\sum_{\xi\in
\mathcal{D}_{d,T}} \frac{A_T}{\tbinom{d+2}{2}}
(f_{\boldsymbol{\Delta}}(\xi)- c_\xi(f_{\boldsymbol{\Delta}}))
c_\xi(g_{\boldsymbol{\Delta}}).
$$
We have
$$
|I_1(T)|= A_T \|\{f_{\boldsymbol{\Delta}}(\xi)- c_\xi(f_{\boldsymbol{\Delta}})\}_{\xi\in \mathcal{D}_{d,T}}\|_\infty  \|\{c_\xi(g_{\boldsymbol{\Delta}})\}_{\xi\in \mathcal{D}_{d,T}}\|_\infty
$$
and hence, by Th. 2.6 in \cite{lai2007spline},
$$
|I_1(T)|\le A_T K^2\|B_d(f_{\boldsymbol{\Delta}}) - f_{\boldsymbol{\Delta}}\|_T
|g_{\boldsymbol{\Delta}}|_T,
$$
where $K$ is a positive constant.
We use the property of Bernstein-B\'ezier approximation again, i.e.
the estimate in (\ref{Lai2.45}) to have
\begin{align*}
|I_1(T)| \le & K^2 A_T \frac{|T|^2}{d} |f_{\boldsymbol{\Delta}}|_{2,T}
\|g_{\boldsymbol{\Delta}}\|_{\infty, \Omega}\cr
\le & K^2 \|f_{\boldsymbol{\Delta}}\|_{2,1,T}
\|g_{\boldsymbol{\Delta}}\|_{\infty, \Omega}.
\end{align*}
Therefore we have
$$
\sum_{T\in \boldsymbol{\Delta}} |I_1(T)|\le K^2 \frac{|T|^2}{d}
\|g_{\boldsymbol{\Delta}}\|_{\infty, \Omega}
\|f_{\boldsymbol{\Delta}}\|_{2,1,\Omega}.
$$
Similarly, we can discuss
$$
I_2(T):=\sum_{\xi\in
\mathcal{D}_{d,T}} \frac{A_T}{\tbinom{d+2}{2}}
f_{\boldsymbol{\Delta}}(\xi ) (c_\xi(g_{\boldsymbol{\Delta}})
 -g_{\boldsymbol{\Delta}}(\xi))
$$
to have a similar estimate as $I_1(T)$.
Putting these three estimates above we
have  obtained
$$|\epsilon_{\boldsymbol{\Delta}}(f_{\boldsymbol{\Delta}},g_{\boldsymbol{\Delta}})|\leq K|\boldsymbol{\Delta}|^{2}
(\|f_{\boldsymbol{\Delta}}\|_{2,2,\Omega}
\|g_{\boldsymbol{\Delta}}\|_{2,2,\Omega} +  \|f_{\boldsymbol{\Delta}}\|_{2,1,\Omega}
\|g_{\boldsymbol{\Delta}}\|_{\infty,\Omega}+
\|f_{\boldsymbol{\Delta}}\|_{\infty,\Omega}
\|g_{\boldsymbol{\Delta}}\|_{2,1,\Omega}),
$$
where $K$ is a positive constant,
$|\boldsymbol{\Delta}|$ is the length of the longest edge in the
triangulation $\boldsymbol{\Delta}$. These complete the proof.
\end{appendix}

\bibliographystyle{Chicago}

\bibliography{references_all}
\end{document}